\documentclass{aa}  

\usepackage{lineno}

\nolinenumbers

\usepackage{txfonts}
\usepackage[hyperindex=true,breaklinks=true,colorlinks=true,linkcolor=blue,filecolor=blue,urlcolor=blue,citecolor=blue,breaklinks=true]{hyperref}
\usepackage{tabularray}
\usepackage{placeins}
\usepackage{float}
\usepackage{soul}
\usepackage{appendix}
\usepackage{multicol}
\usepackage{amsmath}
\usepackage{graphicx}

\begin{document}

   \title{Deriving the star formation histories of galaxies from  spectra with simulation-based inference}

   \titlerunning{Star formation histories from  spectra with simulation-based inference}

   \subtitle{}

   \author{Patricia Iglesias-Navarro \inst{1,2}   \and Marc Huertas-Company \inst{1,2,3,4} \and Ignacio Martín-Navarro \inst{1,2} \and Johan H. Knapen \inst{1,2} \and Emilie Pernet \inst{1,2,5}}

   %check order and more authors

   \institute{Instituto de Astrofísica de Canarias, C/ Vía Láctea s/n, 38205 La Laguna, Tenerife, Spain \and Departamento de Astrofísica, Universidad de La Laguna, 38200 La Laguna, Tenerife, Spain \and
             Observatoire de Paris, LERMA, PSL University, 61 avenue de l'Observatoire, F-75014 Paris, France
          \and  
          Universit\'e Paris-Cit\'e, 5 Rue Thomas Mann, 75014 Paris, France \and
    Department of Physics, Faculty of Engineering and Physical Sciences, University of Surrey, GU2 7XH, Guildford, United Kingdom}

   \date{}

% \abstract{}{}{}{}{} 
% 5 {} token are mandatory
 
  \abstract
   {High-resolution galaxy spectra encode information about the stellar populations within galaxies. The properties of the stars, such as their ages, masses,  and metallicities, provide  insights into the underlying physical processes that drive the growth and transformation of galaxies over cosmic time.
    We explore a simulation-based inference (SBI) workflow to infer  from optical absorption spectra the posterior distributions of metallicities and the star formation histories (SFHs) of galaxies (i.e. the star formation rate as a function of time). 
    We generated a dataset of synthetic spectra to train and test our model using the spectroscopic predictions of the MILES stellar population library and non-parametric SFHs. We reliably estimate the mass assembly of an integrated stellar population with well-calibrated uncertainties. Specifically, we reach a score of $0.97\,R^2$ for the time at which a given galaxy from the test set formed $50\%$ of its stellar mass, obtaining samples of the posteriors in only $10^{-4}$\,s. We then applied the pipeline to real observations of  massive elliptical galaxies, recovering the well-known relationship between the age and the velocity dispersion, and show that the most massive galaxies ($\sigma\sim300$ km/s) built up to 90\% of their total stellar masses within $1$\,Gyr of the Big Bang. The inferred properties also agree with the state-of-the-art inversion codes, but the inference is performed up to five orders of magnitude faster.
    This SBI approach coupled with machine learning and applied to full spectral fitting makes it possible to address large numbers of galaxies while performing a thick sampling of the posteriors. It will allow both the deterministic trends and the inherent uncertainties of the highly degenerated inversion problem to be estimated for large and complex upcoming spectroscopic surveys, such as DESI, WEAVE, or 4MOST.}
    

   \maketitle
%
%-------------------------------------------------------------------

\section{Introduction}

 Understanding the physical processes that regulate star formation over cosmic time is one of the main challenges of galaxy studies, as their evolution depends on a balance between processes that trigger star formation and others that prevent it by expelling or heating gas \citep[e.g.][]{Franx90,Silk_2012, lilly13, nacho}. Reconstructing star formation histories (SFHs) is thus a fundamental step towards understanding galaxy evolution. However, inferring them from observed spectra for a statistically significant sample of galaxies is a complex inverse problem subject to a large number of degeneracies that are not well understood \citep[e.g.][]{worthey92, Ocvirk, Conroy_2009}. 

  Stellar population synthesis (SPS) models are normally used, built from simple stellar population (SSP) templates. SSPs describe the evolution in time of the spectrum of a single stellar burst of fixed metallicity and chemical abundance and require three basic inputs: (i) stellar evolution theory in the form of isochrones, for example Padova \citep{Girardi_2000}, BaSTI \citep{Pietrinferni_2006}, or MIST \citep{Choi_2016},  (ii) empirical or  theoretical stellar spectral libraries, for example CaT \citep{Cenarro_2001}, MILES \citep{Vazdekis_2010}, or XSL \citep{Gonneau_2020}, and (iii) an initial mass function \citep[IMF;][]{salpeter95,Kroupa_2001,Chabrier_2003}. Each of these three inputs can in principle depend on the metallicity and/or elemental abundance patterns.  Composite stellar populations differ from simple ones because they contain stars with a range of ages (as given by their SFHs) and metallicities (as given by their time-dependent metallicity distribution function,  $P$($\rm{[M/H]}$\footnote{i.e. metals over hydrogen, defined as $\rm{[M/H]}=\log \left(\frac{\rm M}{\rm H}\right)-\log \left(\frac{\rm M}{\rm H}\right)_{\odot}$.}, $t$)).\\

To connect a model with observations, a statistical inference framework is required. Full spectral fitting algorithms are a popular tool for inferring stellar population properties from integrated spectra, generally based on a backward comparison between data and models  \citep[e.g.][]{starlight,Ocvirk,vespa,firefly, Cappellari2022}. The success and reliability of this method depend on the quality of the template spectra and the robustness of the fitting algorithm. In parallel, stellar population inversion algorithms have evolved into Bayesian statistics \citep[e.g.][]{Mart_n_Navarro_2019,Johnson_2021,2024maciata, 2024wang}, with the Markov chain Monte Carlo (MCMC) method primarily used for sampling the posterior distributions, which are assumed to be Gaussian, allowing an efficient exploration of the degeneracies associated with the large parameter space.

Existing Bayesian spectral modelling approaches, however, require a substantial computational investment, ranging from $10$ - $100$ CPU hours per galaxy \citep{Carnall_2019,tacchela21}. This computational demand is already high for current datasets, like    the Sloan Digital Sky Survey \citep[SDSS;][]{SDSS} or  the Cosmic Evolution Survey \citep[COSMOS;][]{cosmos}, which contain hundreds of thousands of galaxy spectral energy distributions, and it will become a prohibitive bottleneck for upcoming surveys. Over the next decade, the Dark Energy Spectroscopic Instrument \citep[DESI;][]{desi}, the \textit{William Herschel} Telescope Enhanced Area Velocity Explorer \citep[WEAVE;][]{weave}, the 4-metre Multi-Object Spectroscopic Telescope \citep[4MOST;][]{4most}, and the Multi-Object Optical and Near-infrared Spectrograph \citep[MOONS;][]{MOONS} will capture spectra from billions of galaxies, which would imply tens or hundreds of billions of CPU hours.\\

Machine-learning-based models are becoming more popular in astronomy \citep[e.g.][]{huertascompany2023brief, joanna24,moser24,hunt2024predicting}. In contrast with traditional full-spectral fitting, where SFHs are built by combining SSPs to recover the observed spectra, machine learning models directly learn the implicit relation between the spectra and the SFHs. Its systematic uncertainties are thus different from those from classical spectral fitting, complementing and strengthening the measurements \citep{lovell19}. On the other hand, neural density estimators have recently been introduced to address the computational challenge of Bayesian methods \citep[e.g.][]{kwon24}, demonstrating a significant acceleration in spectral energy distribution  model evaluations of up to five orders of magnitude \citep{Hahn_2022}, so the posterior inference for each galaxy now takes only seconds.\\

In this work we explore a novel approach based on probabilistic machine learning and simulation-based inference (SBI) to estimate the SFHs of galaxies from their optical absorption spectra. We followed the initial steps laid out for this method for full-spectral fitting by \cite{gourav22}.  The main differences from classical Bayesian inference methods are that the SBI approach does not require an explicit functional form for the likelihood, typically considered Gaussian, and that once the model is trained, it can be evaluated on different observations with minimal computational cost. In this paper, which essentially describes a proof of concept,  we do not include noise or emission lines, although we discuss the way to generalise the method in the future to include these features.\\

The outline of the paper is as follows: In Sect.~\ref{methods} we describe the Bayesian inference framework, using SPS (\ref{forward}), an encoder of the spectra (\ref{encoder}),  and a neural density estimator (\ref{normflows}). In Sect.~\ref{results} we test the model with both mocked observations and stacks of early-type galaxies (ETGs) from SDSS. We further discuss the details of the model and the inferred galaxy properties in Sect.~\ref{discussion}.

%-------------------------------------------------------------------
\section{Methods}
\label{methods}

When applied to spectral fitting, Bayesian algorithms aim to infer the posterior probability distributions $P(\theta \mid x)$ of galaxy properties, $\theta$, given observations, $x$. For a specific $\theta$ and $x$, one typically evaluates the posterior using Bayes' rule, $P(\theta \mid x) \propto P(\theta) \; P(x \mid \theta)$, where $P(\theta)$ denotes the prior distribution and $P(x \mid \theta)$ the likelihood. As this last distribution is typically intractable, conventional statistical approaches assume fixed functional forms for it, attempting to be consistent both with the empirical data and the theoretical models.\\

Simulation-based inference, also known as likelihood-free inference, offers an alternative that does not require an explicit form for the likelihood. It consists of creating synthetic data, in the often called the forward model, and then fitting them like real observations, in the backward model, comparing against a known ‘truth’. Therefore, SBI uses a generative model (i.e. a simulation $F$) to generate mock data $x^{\prime}$ given parameters $\theta^{\prime}: F\left(\theta^{\prime}\right)=x^{\prime}$. It uses a large number of simulated pairs $\left(\theta^{\prime}, x^{\prime}\right)$, which can be considered as samples of the implicit likelihood distribution, to directly estimate the posterior $P(\theta \mid x)$, the likelihood $P(x \mid \theta)$, or the joint distribution of the parameters and data $P(\theta, x)$.  This technique has already been successfully applied  to several Bayesian parameter inference problems in astronomy \citep[e.g.][]{Cameron_2012, mishrasharma2022inferring, hahn23,zhang23}.\\

 As most real observations are unbalanced and affected by selection effects, we typically do not have the kind of broad and uniform dataset required by inference algorithms. We can achieve this parameter space coverage with the forward model, although the inference will be more biased in those areas of the parameter space with fewer observations. On the other hand, SBI is especially useful when testing single-effect biases (e.g. parametrisations of SFHs or noise) as it enables complexity to be layered in the simulation. In contrast, it is likely to present problems of domain shifts when generalising results to real observations. Modelling considerations, like priors, wavelength coverage, or resolution, can have a significant impact on inferred galaxy properties \citep{Leja_2019,Hahn_2022} and must be carefully selected.

\subsection{Forward model}
\label{forward}

The first step is to generate a synthetic dataset to train and test our model, for which the spectrum, the SFH, and the metallicity are known. As mentioned before, in this toy model we do not include emission features or noise. We worked with MILES SPS models \citep{Vazdekis_2010}.  These fully empirical SSP templates provide medium resolution \citep[$\text{FWHM}=2.51\AA$,][]{falcon11} predictions for a wide range of ages (from $0.03$ Gyr to $14.00$ Gyr) and metallicities ($\rm{[M/H]}$ moving from $-2.27$ to $0.40$). We selected a Kroupa universal IMF and BaSTI isochrones, in the wavelength range $[3540.5,7409.6] \, \AA$, and used the so-called base models for $[\alpha/\rm{Fe}]$, following the abundance pattern of the solar neighbourhood.\\

We built non-parametric SFHs with Dense Basis \citep{Iyer_2017}, specifically the module GP-SFH. These SFHs allow complex behaviours, like rejuvenation events, bursts, or sudden quenching, and do not rely on a fixed functional form. The selected prior assumes that the fractional specific star formation rate (sSFR; i.e. the star formation rate normalised by the stellar mass), for three equally spaced time bins, follows a Dirichlet distribution with a concentration parameter $\alpha=1$. This prior for the SFHs has been studied in detail in \cite{Leja_2019}. Moreover, a uniform prior is included for the logarithm of the total stellar mass, $8.0 \leq \text{log} \left(M_{*}/M_{\odot}\right) \leq 12.0$, and for the logarithm of the sSFR at $z=0$, $-17.0 \leq \text{log} \left(\text{sSFR}/ \rm{yr}^{-1}\right) \leq -7.5$. We believe that these priors are general enough not to impose strong effects on the inferred galaxy properties; however, they will always play a role in the model estimations.\\

For simplicity, we assumed that there is no time evolution in metallicity, so each SFH was assigned a single value of $\rm{[M/H]}$. We interpolated MILES spectra in time to obtain a SSP spectrum each $0.01347$ Gyr, with cosmic time in the range $[0.00,13.47]$ Gyr ($1{,}000$ time bins). On the other hand, using unequally spaced metallicities imposes a non-uniform prior on $\rm{[M/H]}$, which would bias the inference. To avoid this, we interpolated the spectra in $\rm{[M/H]}$ too, obtaining $15$ equally spaced values of this parameter in the range $[-2.3,0.4]$. For each artificial SFH, given a value of $\rm{[M/H]}$, all the MILES interpolated spectra (corresponding to different ages and with that metallicity) are combined as

\begin{equation}
\resizebox{.49\textwidth}{!}
{
$F_{\text{gal}} \left( \lambda, M_{\text{tot}}, \text{[M/H]}, [\alpha/\text{Fe}]_{\text{Base}} \right) = \sum_{t_i} \frac{M(t_i)}{M_{\text{tot}}} \cdot F_{\text{SSP}} \left( \lambda, t_i, \text{[M/H]}, [\alpha/\text{Fe}]_{\text{Base}} \right)$
}
\end{equation}

\begin{figure*}[h]
    \centering
    \includegraphics[width=0.49\textwidth]{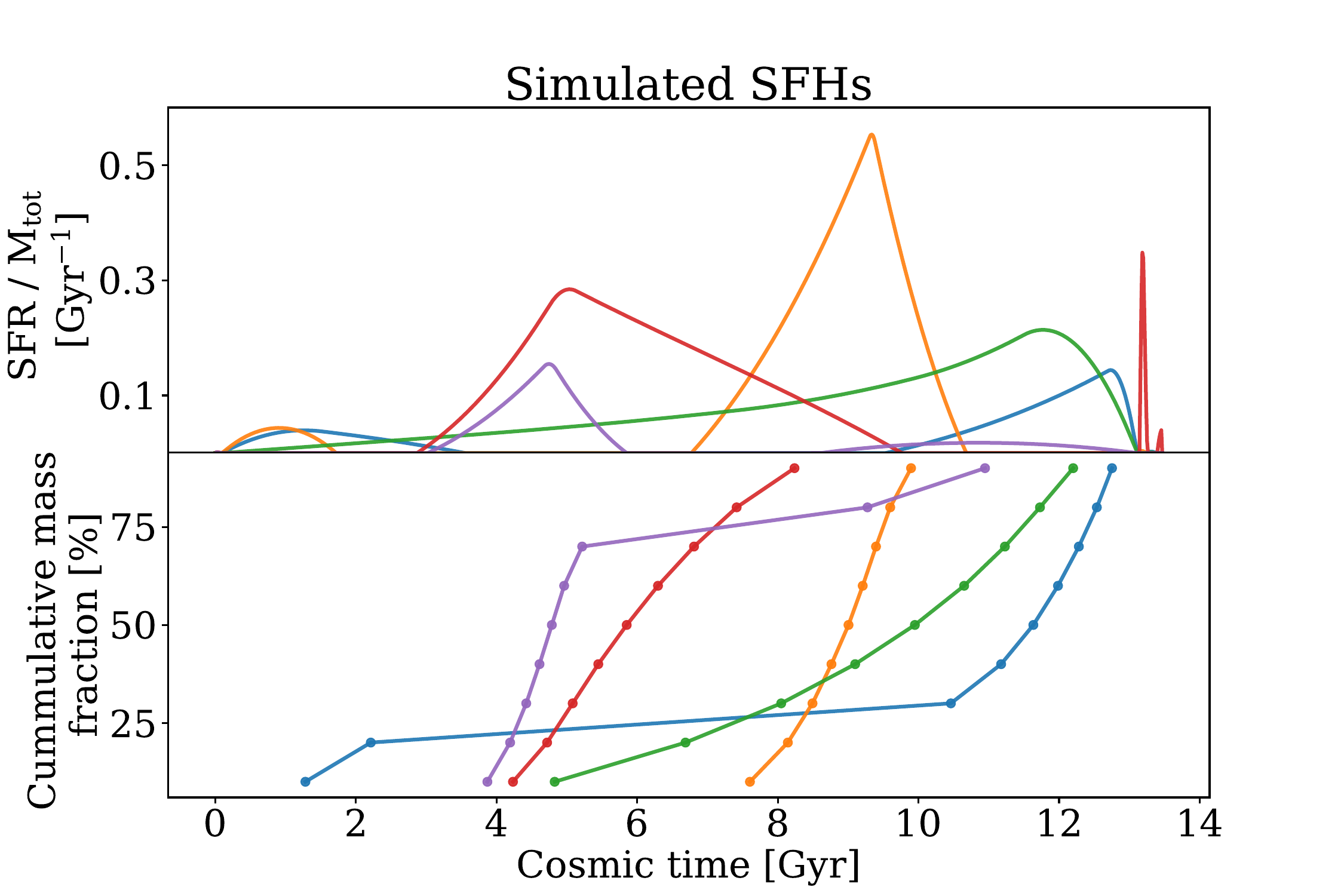}
    \includegraphics[width=0.49\textwidth]{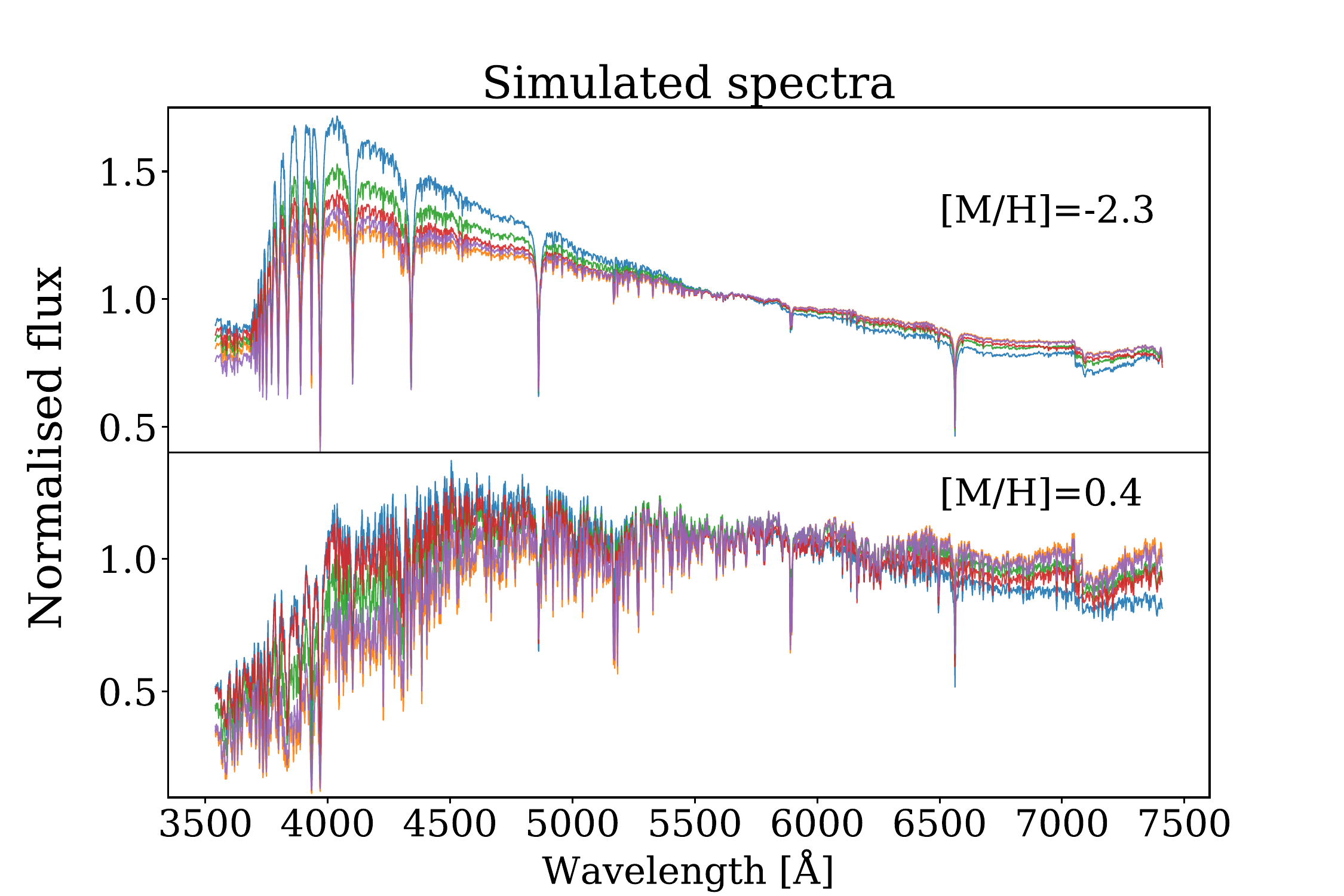}
    \caption{Samples of the simulated dataset, each in a different colour. In the left plot we show their mocked SFHs: in the upper panel the star formation rate normalised by the total stellar mass in Gyr$^{-1}$ and in the lower one the nine stellar mass percentiles, together with three dashed grey lines indicating when $25\%$, $50\%,$ and $75\%$ of the total stellar mass has been formed. In the right plot, we show the  simulated normalised spectra corresponding to these SFHs (same colours) for two different values of metallicity: in the upper panel we fix $\rm{[M/H]}=-2.3$, and in the lower one  $\rm{[M/H]}=0.4$, which are respectively the minimum and maximum bins of $\rm{[M/H]}$ in the simulation.}
    \label{input}
\end{figure*}

Finally, each spectrum was normalised by its median. In total, we obtained $10{,}000$ SFHs for each value of $\rm{[M/H]}$, so: $10{,}000$ SFHs x $15$ bins of metallicity $= 150{,}000$ spectra, for which SFHs and metallicities are known. Examples of the mock SFHs and their corresponding spectra are shown in Fig.~\ref{input}.\\

 Stellar mass curves (i.e. the star formation rate as a function of time) are integrated to get nine stellar mass percentiles. We thus focused on the cosmic time (the time since the Big Bang) at which $10$\%, $20$\%, ... $90$\%  of the total stellar mass of the galaxy has been formed. These quantities are more robust than their non-cumulative analogues, as only the features in the SFHs that contribute meaningfully to the total stellar mass of the galaxy play a role. This approach decreases the dimensionality of the SFHs without relying on standard parametric shapes, and mitigates the impact of possible artefacts in the synthetic SFHs.

\subsection{Backward model}
\label{backward}
\subsubsection{Encoding the spectra}
\label{encoder}

High-quality absorption spectra typically cover at least $1000 \, \AA$, described by a similar number of pixels. However, the widespread adoption of template libraries suggests that galaxy spectra in fact occupy a low-dimensional manifold. In particular, \cite{Portillo_2020} demonstrated that a high-fidelity reconstruction can be achieved by an autoencoder (AE) architecture  \citep{hinton}, a non-linear dimensionality reduction technique, with a latent space of just six dimensions. \cite{Teimoorinia_2022}  and  \cite{melchior2022} improved the method by introducing convolutional elements into the AE to aid the extraction of correlated features from the spectra. In short, AEs are feed-forward neural networks that learn efficient encodings of data in an unsupervised manner. They consist of two parts: an encoder, which takes data as input and compresses it to produce a low-dimensional latent representation, and a decoder, which takes the latent representations and decompresses them to reconstruct the original data. Due to their non-linear behaviour, AEs can capture non-linear features, such as line widths, with fewer parameters than principal component analysis \citep[][]{kramer91,Portillo_2020}, one of the most commonly used techniques. Moreover, unlike line-ratio diagnosis, AEs use the continuum information in the spectra, resulting in an interpretable latent space where galaxies of the same type are grouped together, even though the AEs were never given these classifications in training.\\

We took advantage of this approach to obtain low-dimensional representations of the spectra, using the encoder part of the architecture implemented by \cite{melchior2022} (\href{https://github.com/pmelchior/spender.git}{SPENDER\footnote{\href{https://github.com/pmelchior/spender.git}{https://github.com/pmelchior/spender.git}}}). These representations are more suitable than the full spectrum to be introduced into a Bayesian inference model for a fully probabilistic treatment, as they provide summary statistics of the spectra by reducing the dimensionality and disentangling correlations. The size of the latent representations is a tunable parameter and depends on what information in the spectrum is relevant to determine the SFHs and metallicity, as well as on the dimension of the spectrum and the performance required. During this work, it is set to $16$, as most of the details of the spectra that contain information relevant to the early star formation of galaxies are extremely subtle.\\

 Our network consists of a three-layer convolutional neural network (moving to wider kernels and including max-pooling), plus an attention module (dot-product) and a three-layer multi-layer perceptron to obtain the latent vectors. Furthermore, an additional two-layer multi-layer perceptron has been included to optimise the encoding for our final task: to obtain stellar mass percentiles and a value of $\rm{[M/H]}$. We thus incorporated a log-cosh loss function\footnote{$L =\frac{1}{n} \sum_{i=1}^n \log \left(\cosh \left(\hat{\theta}_i-\theta_i\right)\right)$} that computes the difference between the predicted and real values, driving the evolution of the training. We selected this loss function  because it combines the robustness to outliers of the mean absolute error function and the smoothness and differentiability of the mean squared error function \citep{logcosh}.\\
 
The encoder was trained with the training and validation sets ($80\%$ and $10\%$ of the total number of samples), and the training was stopped when the validation loss reached a plateau, after 250 epochs approximately. Then, the model was applied to test data (the remaining $10\%$ of the total number of samples, never seen before by the network), producing not only the latent vectors we used as summary statistics for the Bayesian inference, but also an initial prediction for the mass percentiles and metallicity. These predictions will not be used once we have trained the encoder, as we only require them to drive the training process towards latent representations that preserve the information of these properties, while their probabilistic inference will be done by the neural density estimator, as detailed below.

\subsubsection{Posterior estimation}
\label{normflows}

We used a neural density estimator known as  Normalising Flows to estimate the posterior probability distribution for each of the stellar mass percentiles, and for the metallicity. In a few words, variables $(z)$ described by a simple base distribution $P(z)$, such as a multivariate Gaussian, are transformed through a parameterised invertible transformation $x=f(z)$, which has a tractable Jacobian. The target density $P_{\rm{f}}(x)$ is then given by the change-of-variables formula as a product of the base density and the determinant of the transformation's Jacobian:

\begin{equation}
P_{\rm{f}}(x)=P(z)\left|\operatorname{det} J_{f^{-1}(x)}\right|.
\end{equation}

Several such steps can be stacked, with the probability density `flowing' through the successive variable transformations. The parameters of the transformations during the training of the model are estimated by maximising the log likelihood of the transformed data under the Gaussian distribution $P(z)$, which is indeed tractable:

\begin{equation}
\log P_{\rm{f}}(x)=\log P(f^{-1}(x))+\log \left|\operatorname{det} J_{f^{-1}(x)}\right|.
\end{equation}

Otherwise, it would not be possible to compute the loss since $P(x)$ is unknown. Following this pipeline, Normalising Flows have been generalised to model a conditional density such as  the posterior $P(\theta \mid x)$. Following the choice of \cite{Hahn_2022}, we developed a masked autoregressive flow \citep[MAF;][]{papamakarios2018masked}, implemented with the module SBI \citep{tejerocantero2020sbi},  with five Masked Autoencoder for Distribution Estimation (MADE) blocks, each of them with two hidden layers of $128$ hidden units. In total, the model has a set of $50{,}560$ free parameters, $\phi$. Our goal is to determine the $\phi$ for the MAF model, so that given galaxy properties $\theta$ and summary statistics of the observations $x$, $P_{\phi}\left(\theta \mid x \right)$ accurately estimates the posterior probability distribution $P\left(\theta \mid x \right)$.  In practice, we divided the training data into training and validation sets with a $90/10$ split. We used the Adam optimiser with a learning rate of $5\cdot 10^{-4}$. To prevent overfitting, we evaluated the likelihood of the data points under the base distribution, using the validation data at every training epoch, and stopping the training when that validation likelihood fails to increase after $20$ epochs.\\

 By using this workflow, we trained our model with mock synthetic data and then applied it to real observations to obtain promptly posteriors for the SFHs and metallicity, which was possible thanks to its amortised nature (i.e. it is not focused on any particular observation, as once we have a trained model, new data can be evaluated without repeating the training, the computationally expensive step). We are making two key assumptions. First, the simulator $F$ must generate mock data, $x'$, that are nearly identical to the real data, $x$. Therefore, we should include all possible features one can find in spectra, such as noise or outliers, in the forward model. Second, the neural density estimator must well trained, so  $P_{\rm{f}}(\theta \mid x')$ provides a reliable approximation of $P(\theta \mid x')$, and consequently of $P(\theta \mid x)$. As opposed to standard variational inference, we are not assuming any functional form for the posterior, and since neural networks are universal approximators, we could therefore estimate the posterior with an arbitrarily small error.

%--------------------------------------------------------------------

\section{Results}
\label{results}

The results are presented as follows. First, in Sect.~\ref{post},  the performance of the model is tested by  comparing the derived posteriors with the true values of the physical properties. We compute the time it takes for the full model (encoder and neural density estimator) to obtain the distributions given one spectrum. We also check the uncertainty estimation with a test of statistical coverage \citep{talts2020validating} in Sect.~\ref{sbc}. Finally, we further validate our model in Sect.~\ref{obs} by inferring the SFHs and the metallicity for 18 stacks of ETGs \citep{La_Barbera_2013}, and in Sect.~\ref{compare_ppxf} by comparing them with the results obtained through a consolidated spectral fitting method \citep{Cappellari2022}.

\subsection{Recovering the properties of synthetic galaxies}
\label{post}

We encoded the spectra to obtain 16-component latent representations, optimised to introduce them in the Bayesian inference model by preserving the relevant information to  determine the SFHs and metallicity. Once the encoder is trained, the latent representations for the full dataset are obtained. We study in detail the structure of this latent space in Appendix \ref{encoding}.\\

Then, the neural density estimator is trained with $90$\% of the generated samples ($x=$ {latent vectors}, $\theta=$ {nine stellar mass percentiles, $\rm{[M/H]}$}), with $10\%$ of these composing the validation set. Once the training is finished ($\sim4$ hours: $1$ hour for the encoder and $3$ hours for the Normalising Flows), the remaining $10$\% of the samples is used to test its performance, by obtaining probability distributions for the values of each percentile and $\rm{[M/H]}$ from their latent representations, and comparing the distributions with the true values. Each posterior estimation for the test set is performed with $1{,}000$ samples, taking $\sim0.4$\,s to get the predictions for the $10$ quantities of each galaxy.

\noindent All the time estimations have been made using a NVIDIA Tesla P100 PCIe GPU with 12GB.\\

\begin{figure}[t]
    \centering
    \includegraphics[width=0.49\textwidth]{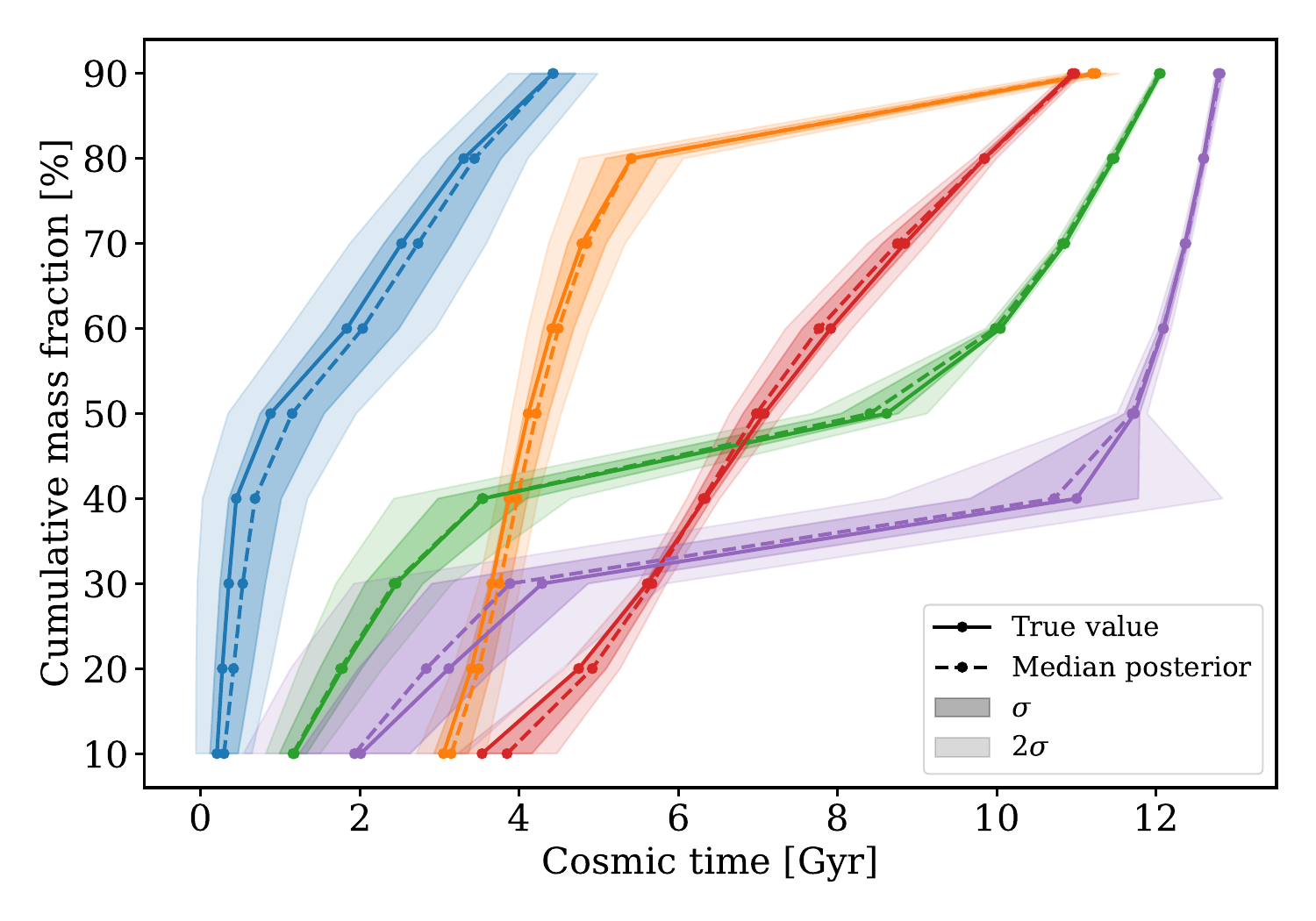}
    \caption{Percentile predictions for five synthetic galaxies. The cumulative mass curves indicate the time at which the nine stellar mass percentiles are achieved over cosmic time. The solid lines correspond to the true values, and the dashed ones to the predictions (medians of the posterior distributions). The $\sigma$ and $2 \sigma$ intervals of confidence are shaded dark and light, respectively. The model yields reliable reconstructions for all five galaxies.}
    \label{examples}
\end{figure}

\begin{figure}[h!]
    \centering
    
    \includegraphics[width=0.285\textwidth]{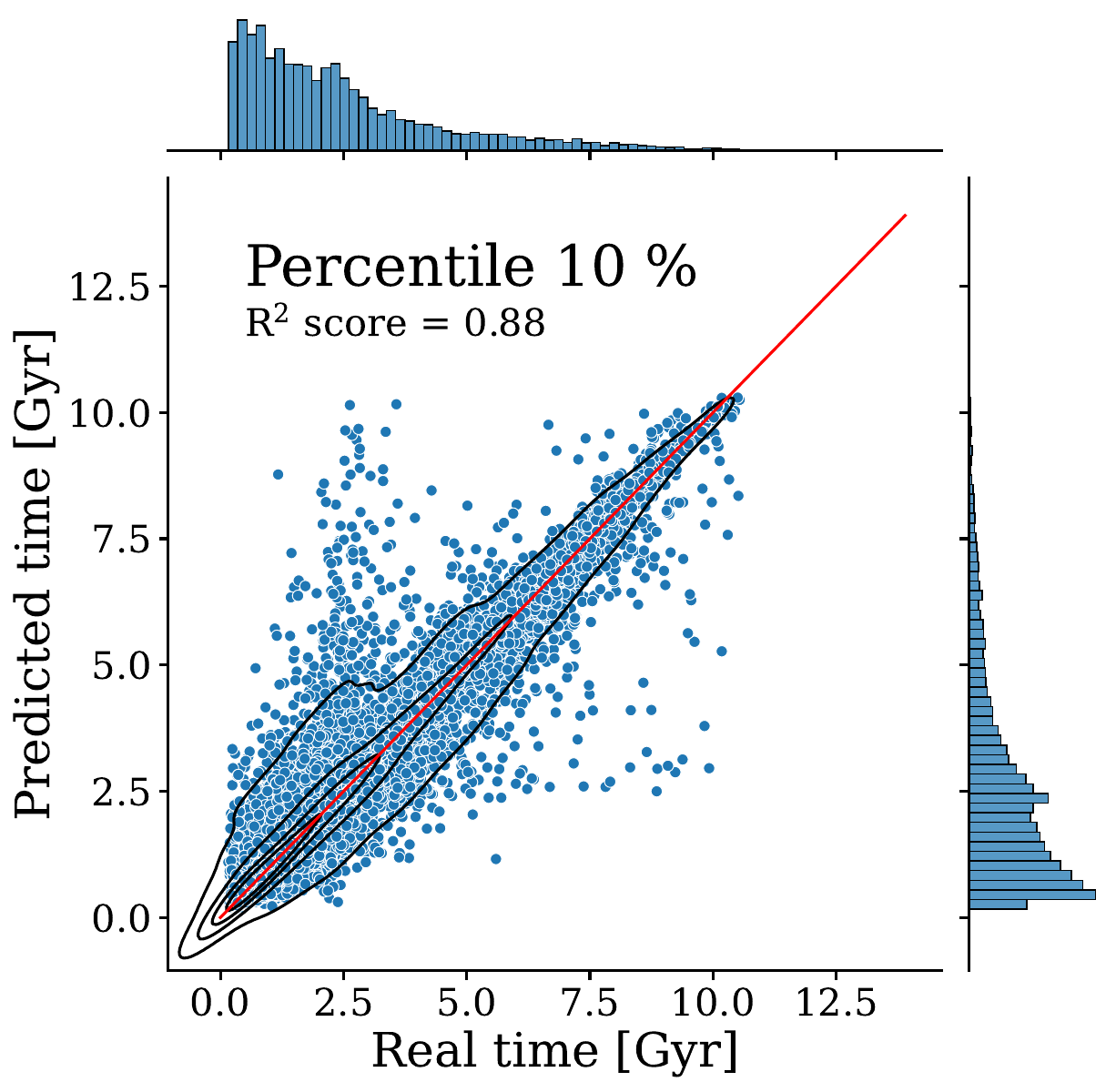}
    \includegraphics[width=0.285\textwidth]{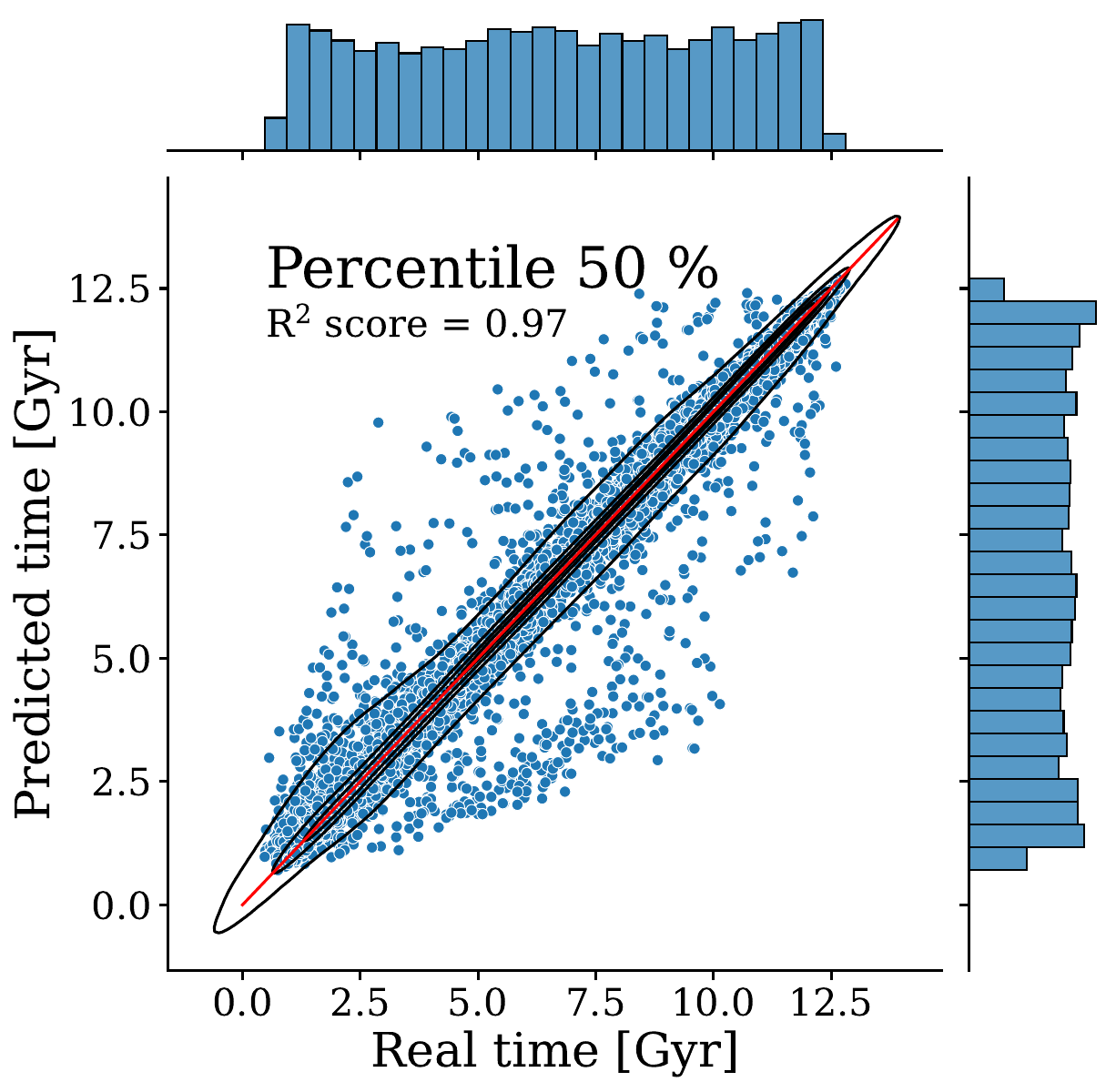}
    \includegraphics[width=0.285\textwidth]{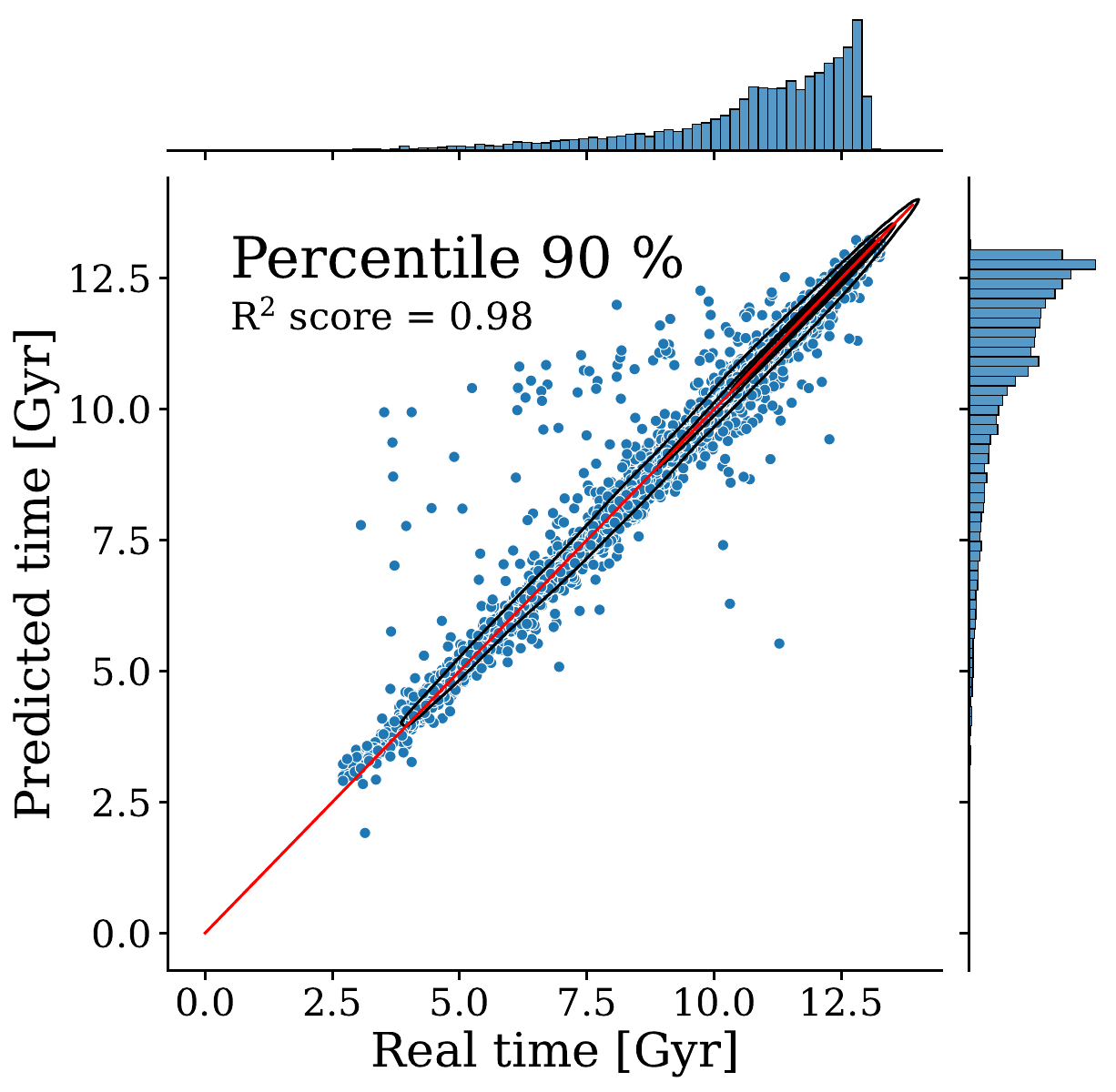}
    \includegraphics[width=0.285\textwidth]{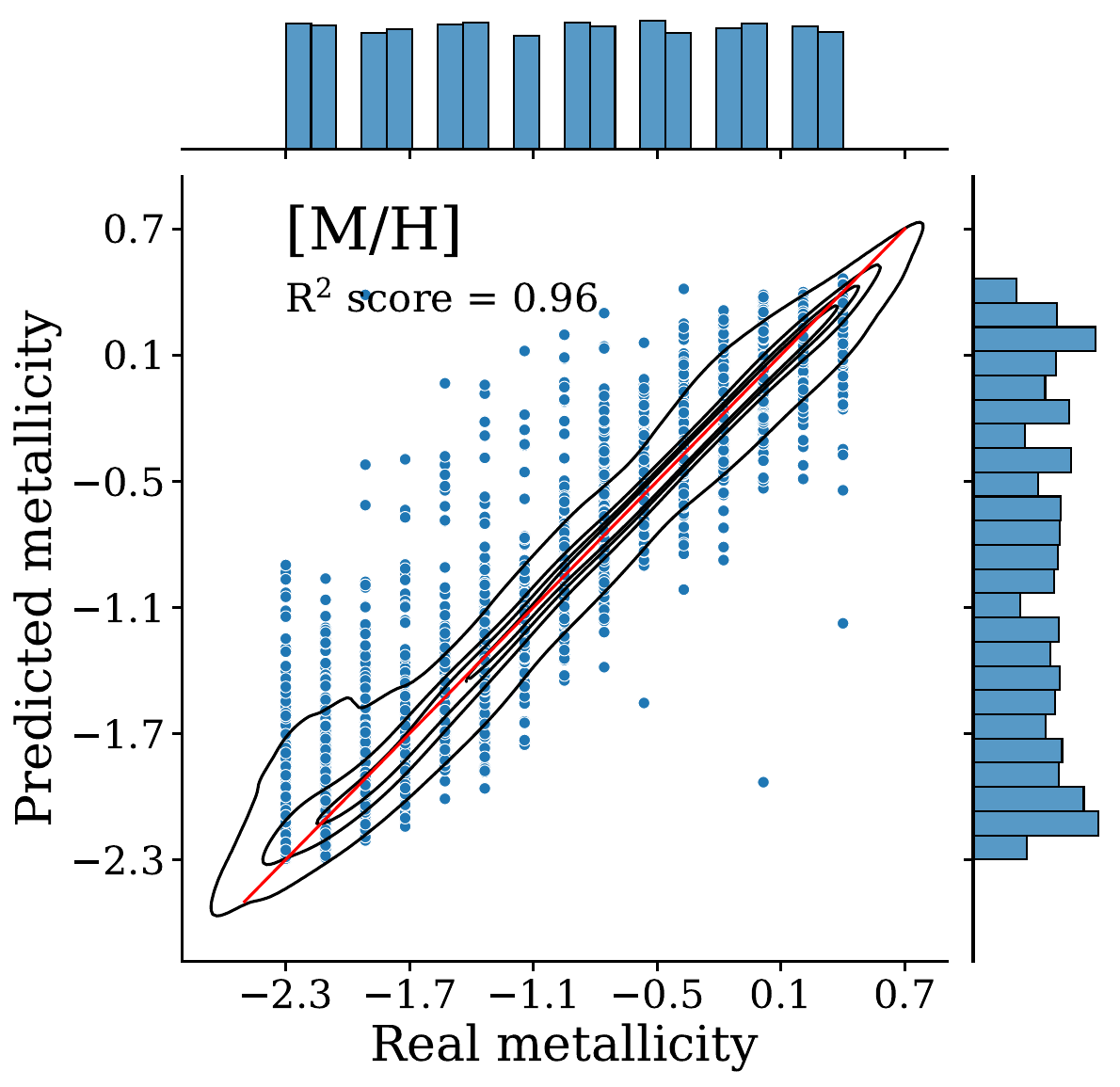}
    
    \caption{Median values of the posterior distributions estimated for the percentiles $10\%$, $50\%$, and $90\%$ and $\rm{[M/H]}$, compared to the true values. The $R^2$ score achieved for each prediction is $0.88$, $0.97$, $0.98$, and $0.96$, respectively. Each blue dot is a different sample from the test set. The red line shows the one-to-one relation, the histograms at the right of each panel show the marginal distributions of the predictions, and the histograms of the real data are shown at the top. Kernel density estimation  contours are drawn in black at iso-proportions of the density of samples.}
    \label{meanvstrue}
\end{figure}

Figure \ref{examples} shows five examples of synthetic SFHs from the test set, presenting the true values for the mass percentiles, the recovered medians of the posteriors, and the 1 and 2$\sigma$ confidence intervals. It its clear that the model can recover from the latent vectors SFHs that are fully consistent with the ground truth within the expected uncertainties. In Fig.~\ref{meanvstrue}, we plot the median values of the posterior distributions predicted for the $15{,}000$ test galaxies against the true values, for the percentiles $10\%$, $50\%$, and $90\%$ (in cosmic time), and for the metallicity. Both agree, close to the one-to-one relation, reaching high $R^{2}$ values\footnote{Also known as coefficient of determination $\displaystyle R^2=1-\frac{\sum_{i=1}^n\left(\theta_i-\widehat{\theta}_i\right)^2}{\sum_{i=1}^n\left(\theta_i-\overline{\theta}_i\right)^2}$, not to be confused with the square of the Pearson's coefficient.} of $0.88$, $0.97$, $0.98$, and $0.96$, respectively.  A larger scatter is observed for earlier percentiles, which is expected as the luminosity of young stars `outshines' the spectra, hiding the information of the oldest ones. This is properly captured by the width of the posteriors, with systematically higher uncertainties for earlier percentiles.\\

\begin{figure}[h!]
    \centering
    
    \includegraphics[width=0.5\textwidth]{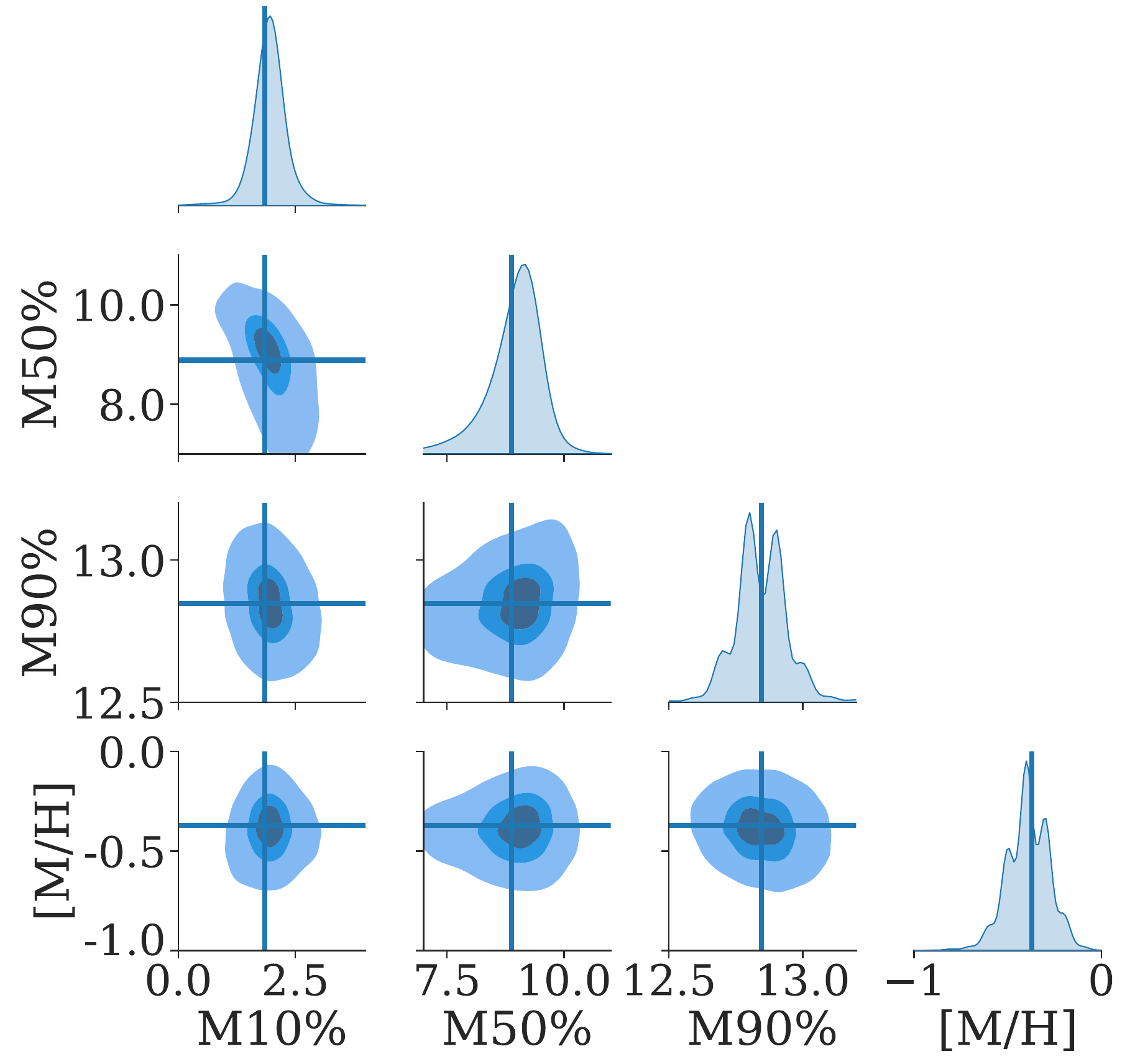}

    \caption{Corner plot with the posterior distributions for a simulated galaxy, drawn with $10{,}000$ samples for the percentiles $10\%$, $50\%$, and $90\%$ (in Gyr) and for $\rm{[M/H]}$. Kernel density estimation contours are shown with different shades at iso-proportions of the density of samples. The solid lines correspond to the true values.  The distributions deviate slightly from Gaussian functions, showing  multi-modalities associated with the degeneracy of the parameter space, but are in perfect agreement with the true values. }
    \label{corner}
\end{figure}

Posterior distributions for the percentiles $10\%$, $50\%$, and $90\%$, and for the metallicity are shown as a corner plot in Fig.~\ref{corner}, for a single galaxy from the test set. These have been computed with $10{,}000$ samples. They deviate slightly from Gaussian functions, showing multi-modalities associated with the degeneracy between the age (in particular the percentile $90\%$)  and  the metallicity, but in perfect agreement with the true values. Figures \ref{examples}, \ref{meanvstrue}, and \ref{corner} confirm that the model is indeed capable of recovering the stellar mass growth and metallicity, with uncertainties associated with the complexity of the inversion problem, due to the very nature of the spectra and to the observational imprints left by galaxies on them.

\subsection{Simulation-based calibration}
\label{sbc}
A key point is ensuring that the estimated posterior distributions are well calibrated. We used simulation-based calibration (SBC) as described in  \cite{talts2020validating} to examine the distribution of the rank statistics of the true parameter values within the marginalised posteriors, which must be uniform if the posterior samples are consistent with the prior used to generate the spectra in the forward model. The only requirement of this test is that we have a generative model for our data, so that we are able to simulate observations from the physical parameters we inferred. However, the original simulation includes more free parameters than the stellar mass percentiles and metallicity, as we used $1,000$ points of the SFHs (i.e. a linear combination of $1,000$ SSP spectra). To carry out this test, we repeated the model training  with a simulation that only takes the nine percentiles and metallicities into consideration, performing the linear combination with nine spectra so that with the posteriors provided by the model we are able to recover unequivocally the synthetic observations.\\

\begin{figure*}[h!]
    \centering
    
    \includegraphics[width=0.45\textwidth]{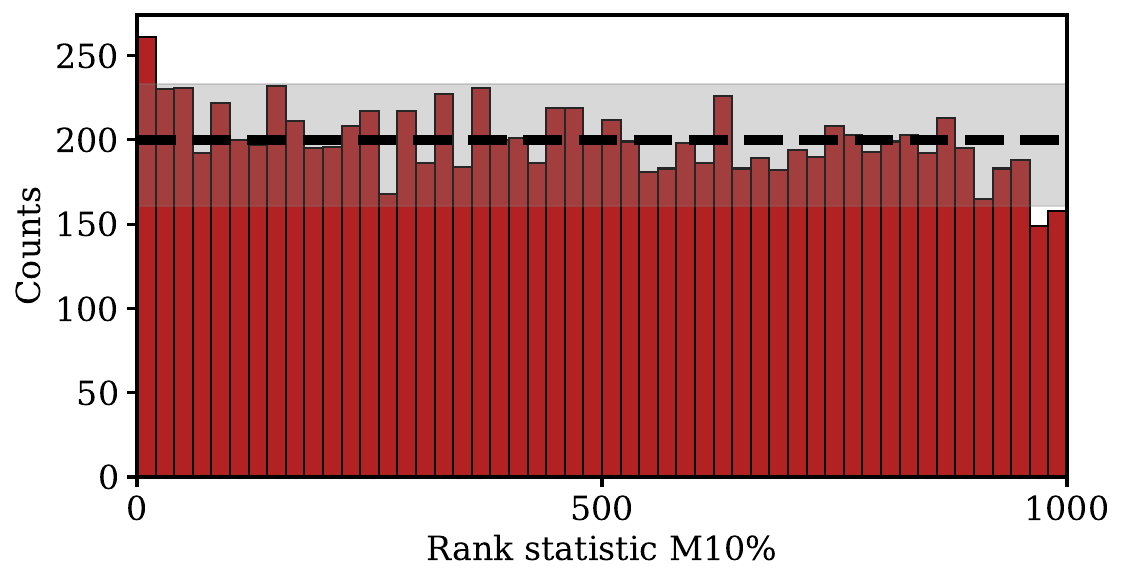}
    \includegraphics[width=0.45\textwidth]{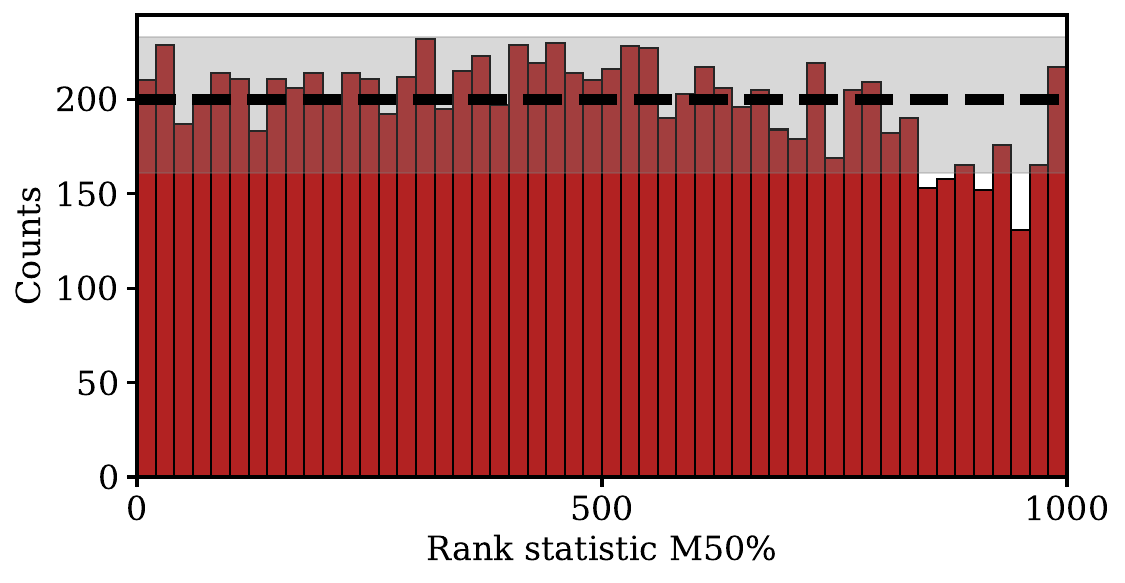}
    \includegraphics[width=0.45\textwidth]{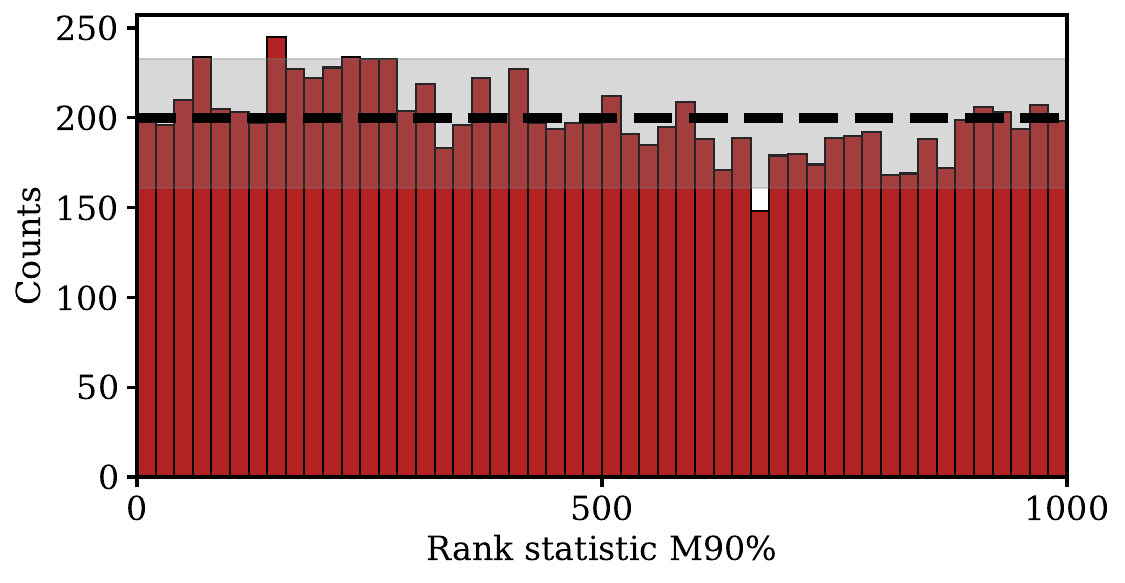}
    \includegraphics[width=0.45\textwidth]{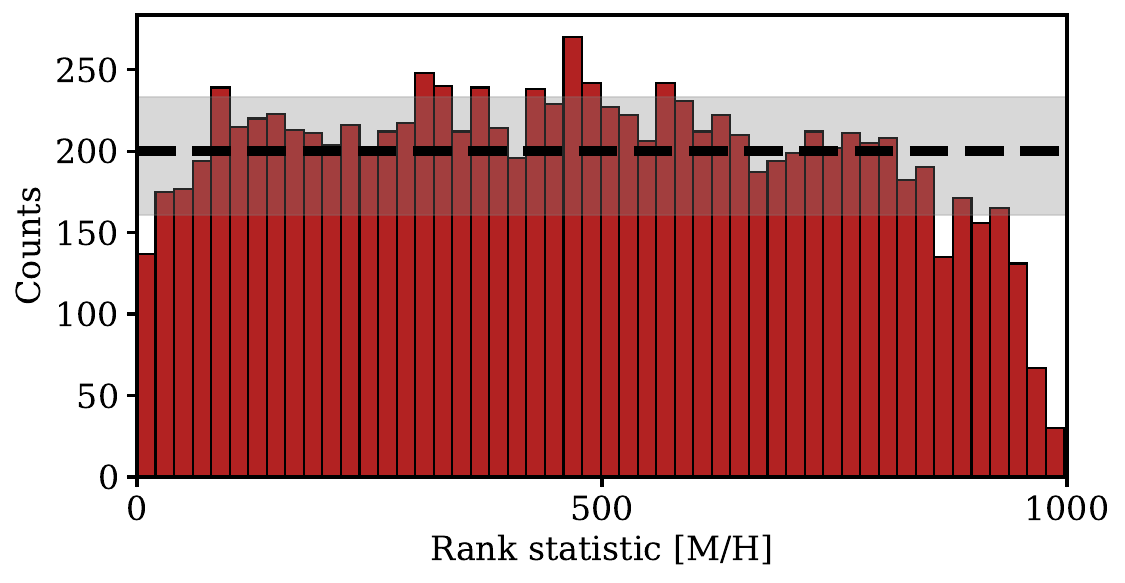}

    \caption{SBC test of the posteriors for $1{,}000$ synthetic test observations. The red histograms in each panel represent the distribution of the rank statistic of the true value within the marginalised posterior for the percentiles $10\%$, $50\%$, and $90\%$ and for $\rm{[M/H]}$. For a well-calibrated posterior, the rank statistics will have a uniform distribution (dashed black line). The grey band indicates $99\%$ of the variation expected from a uniform histogram. The rank statistic distributions of our posteriors are nearly uniform for all four parameters, and therefore, the model provides unbiased and accurate posteriors.}

    \label{ranks}
\end{figure*}

\begin{figure}[h!]
    \centering
    
    \includegraphics[width=0.5\textwidth]{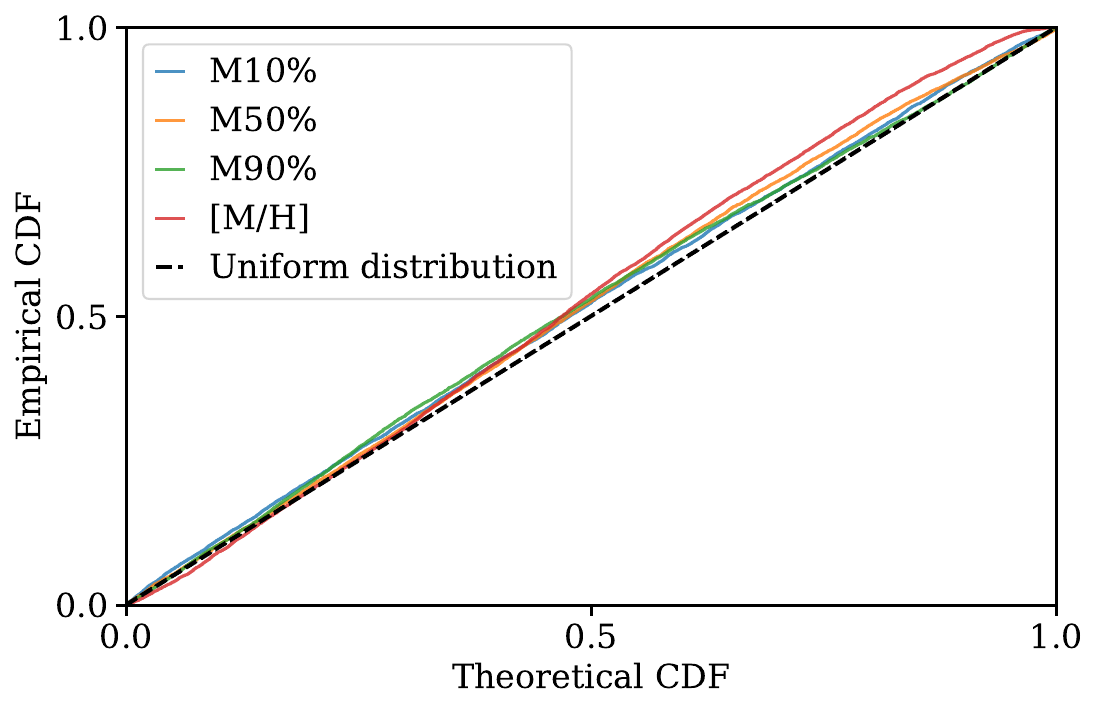}
    
    \caption{Empirical CDFs for the percentiles $10\%$, $50\%$, and $90\%$ and for the $\rm{[M/H]}$ obtained with $1{,}000$ synthetic test observations.  If the posteriors are well calibrated, the nominal coverage probability (the fraction of the probability volume), on the $x$-axis, will be equal to the coverage probability (the fraction of actual values in such a volume), on the $y$-axis, making the CDF diagonal (dashed black line). Our CDFs closely follow the diagonal, indicating that the posteriors are properly computed.}

    \label{ecdf}
\end{figure}

Figure \ref{ranks} shows a histogram from an ensemble of rank statistics of $1{,}000$ prior samples relative to the corresponding posterior samples,  for the percentiles $10\%$, $50\%$, and $90\%$, and for $\rm{[M/H]}$. Each histogram is complemented with a grey band indicating $99\%$ of the variation expected from a uniform histogram. Additionally, in Fig.~\ref{ecdf} we include the empirical cumulative distribution function (CDF), together with the diagonal expected from a well-calibrated posterior. Both plots show uniformly distributed ranks, and consequently an optimal overall performance, except for the metallicity, where we detect a slight $\cap$ shape. This symmetric deviation, related with the wide binning of $\Delta\rm{[M/H]}=0.1929$ we used throughout the work\footnote{A finer binning is always possible, but at the cost of a larger training set and longer training time.}, implies that the computed posterior will be wider than the true posterior, allowing us to estimate an upper limit for the uncertainty of the parameter. This is strictly true in the simulated dataset, but not in real observations with finite S/N, where the posteriors may underestimate the uncertainty as we have not taken the observational noise into account.

\subsection{Testing with observations}

\label{obs}

To test the method, we inferred the SFHs and metallicity for $18$  stacks of ETGs at $z \sim 0$ from SDSS spectra (with velocity dispersion in the range $100-320$ km/s), whose stellar populations have been studied in detail in \cite{La_Barbera_2013}. We highlight that the main advantage of working with these stacks is their very high S/N, and the absence of emission features, which is important because no noise models or emission lines have been incorporated in the simulation.\\

First, we convolved all the observations to emulate the maximum velocity dispersion of the dataset. Then, we clipped the wavelength range to $[4023,6000]$ $\AA$, the maximum window that is common to all stacks. We also processed the MILES spectra used for training to simulate the conditions of the observations. Once all the artificial spectra had the same resolution and wavelength range as the observations, we repeated the training and then tested the model again, analysing the impact of these changes on the performance.\\

As a result of the processing, the $R^2$ score on the synthetic test set decreases to $0.72$, $0.95,$ and $0.96$ for the estimation of the percentiles $10\%$, $50\%,$ and $90\%$, respectively, and to $0.96$ for the metallicity. These losses in performance, mainly for the first percentiles, are a direct consequence of reducing the resolution in the spectral lines, and clipping the Balmer jump in the bluer region of the spectra.\\

\begin{figure}[h!]
    \centering
    \includegraphics[width=0.5\textwidth]{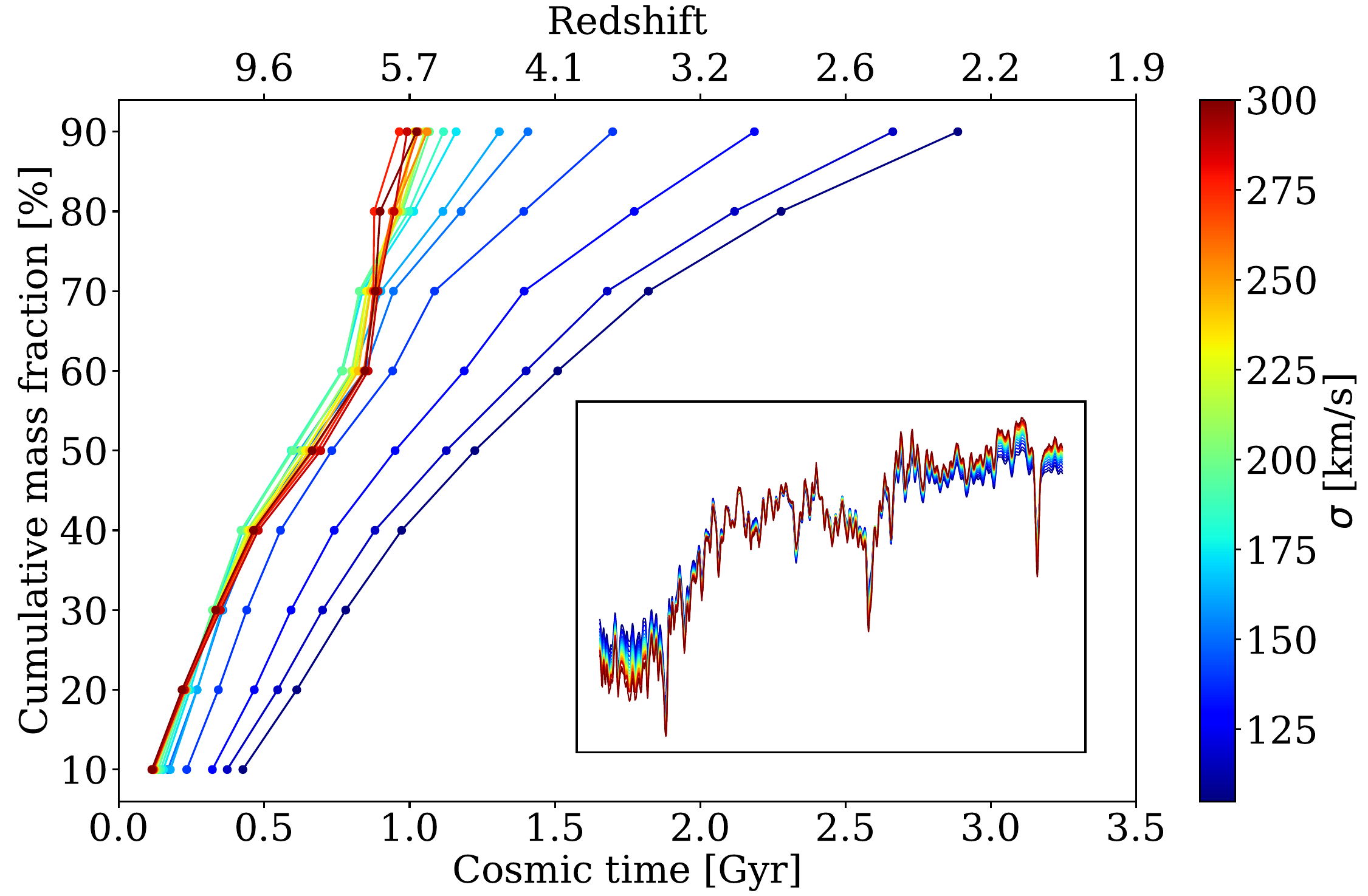}
    \caption{Median values of the posterior distributions obtained for the stellar mass percentiles of $18$ stacks, as a function of cosmic time (in Gyr) and redshift, coloured according to their velocity dispersions (in km/s).  Redder colours indicate higher velocity dispersions (more massive galaxies), and bluer ones lower velocity dispersions (less massive ones). In the inset, we plot the observed spectra, normalised and in the wavelength range $[4023, 6000] \; \AA$, with the same colour map.}
    \label{percentiles_obs_full}
\end{figure}

We then applied the trained model to the spectra of the stacks, generating $10{,}000$ samples from the posteriors. In Fig.~\ref{percentiles_obs_full}, we show the median values of the inferred distributions for the mass percentiles, as a function of cosmic time and redshift\footnote{Assuming a Planck13 cosmology \citep{planck13}.}, using a colour map based on the velocity dispersions. The most massive galaxies (highest velocity dispersions) build up their stellar masses more abruptly, up to 90\% of their total stellar mass by $1$ Gyr after the Big Bang, while the growth of stellar mass is softened as we move to less massive ones.\\

For a more detailed analysis, the median values of the posterior predictions, together with their uncertainties given by one standard deviation, are included in Table~\ref{table_obs}, for all $18$ observations. The quantities shown are the times by which $10$\%, $50$\%, and $90$\% of the total stellar mass have been formed, as well as the metallicity. We do not find a clear trend between the metallicity and the velocity dispersion, but the margin of uncertainty is too large to make any further assumptions, mainly due to the wide binning in $\rm{[M/H]}$ of $0.1929$\,dex for the simulated sample. This is consistent with the results of \cite{La_Barbera_2013}, who reported a variation of $\sim 0.2$\,dex in metallicity from the least massive to the most massive stack.\\

Finally, we carried out a spectral reconstruction with our model in Fig~\ref{spectra_obs}, for the stacks with velocity dispersions $105$, $205,$ and $300$ km/s. This test is usually referred to as posterior predictive check (\citealp{talts2020validating}), and allows us to verify that the properties we infer are indeed compatible with the observed spectra. These observed and reconstructed spectra must be close to each other in the latent space, but a priori this proximity is not trivial in the wavelength space, since we did not minimise the differences with respect to the observed spectra as in more traditional approaches, but performed the Bayesian inference on the latent representations. Even so, it is the most direct way we have of determining the reliability of the inference, and we did it by performing a linear combination of nine SSP MILES spectra of ages obtained from the percentiles of stellar mass, and metallicity fixed and equal to the predicted value. According to the definition of the percentiles, all nine spectra are weighted with $1/9$. The results are very positive, showing mean residuals of  $1.7\%$ between the fit and the observations, averaging over all $18$  stacks and wavelengths. The uncertainties are propagated into the spectra, within the grey-shadowed stripe in Fig~\ref{spectra_obs} corresponding to the interval of two standard deviations.

\subsection{Comparison with existing codes}
%to complete
\label{compare_ppxf}

\begin{figure}[h!]
    \centering
    \includegraphics[width=0.5\textwidth]{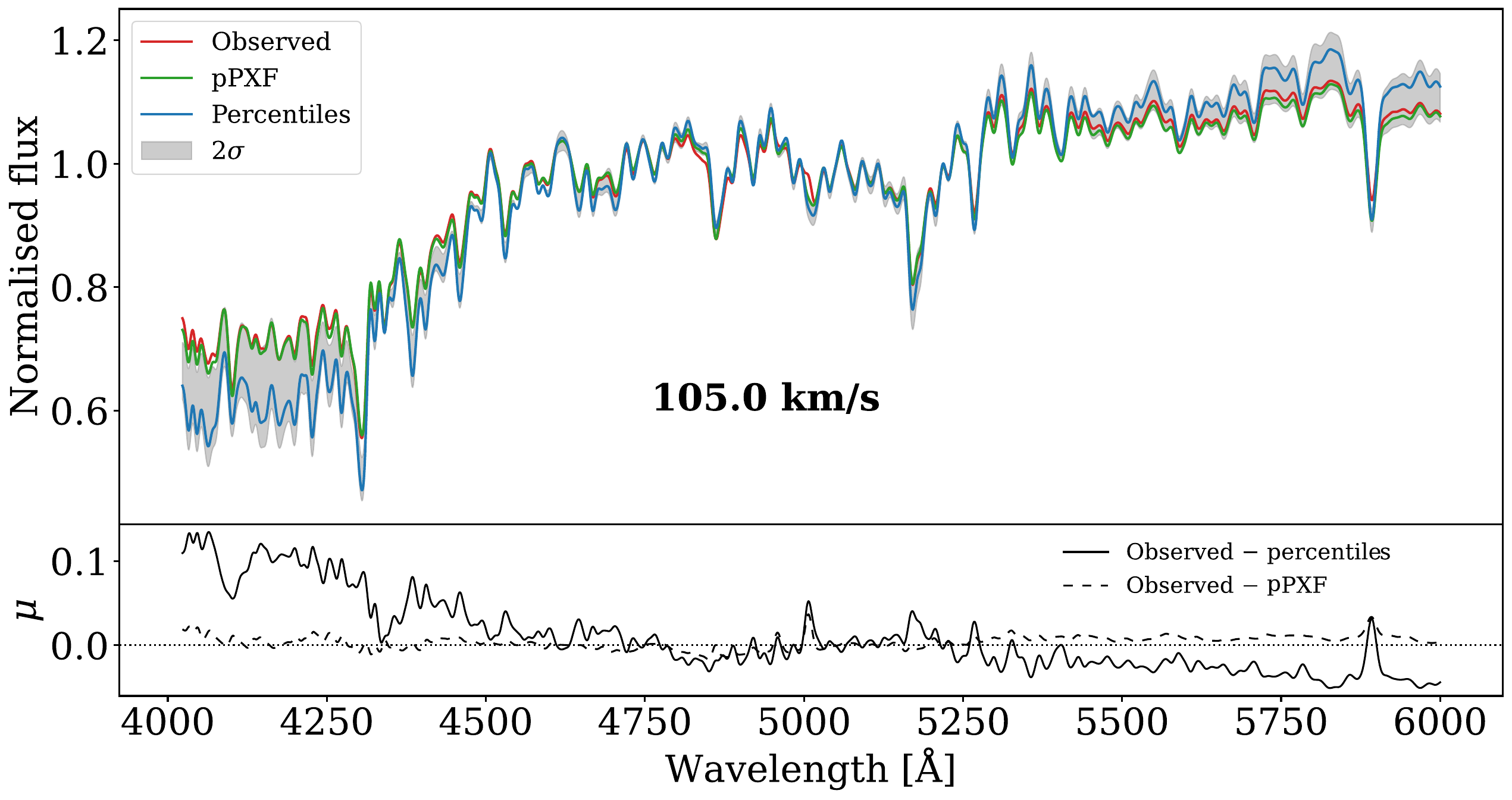}
    \includegraphics[width=0.5\textwidth]{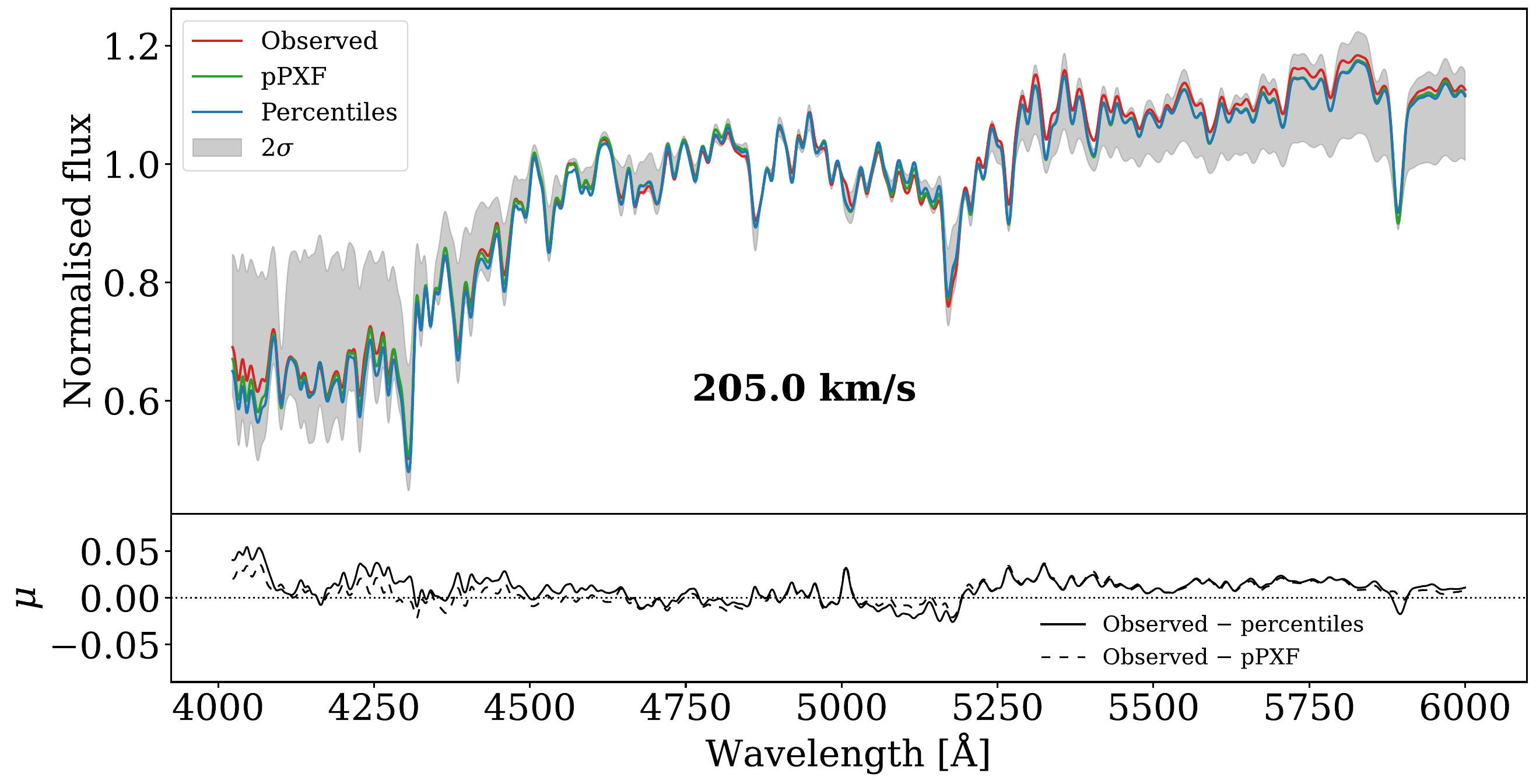}
    \includegraphics[width=0.5\textwidth]{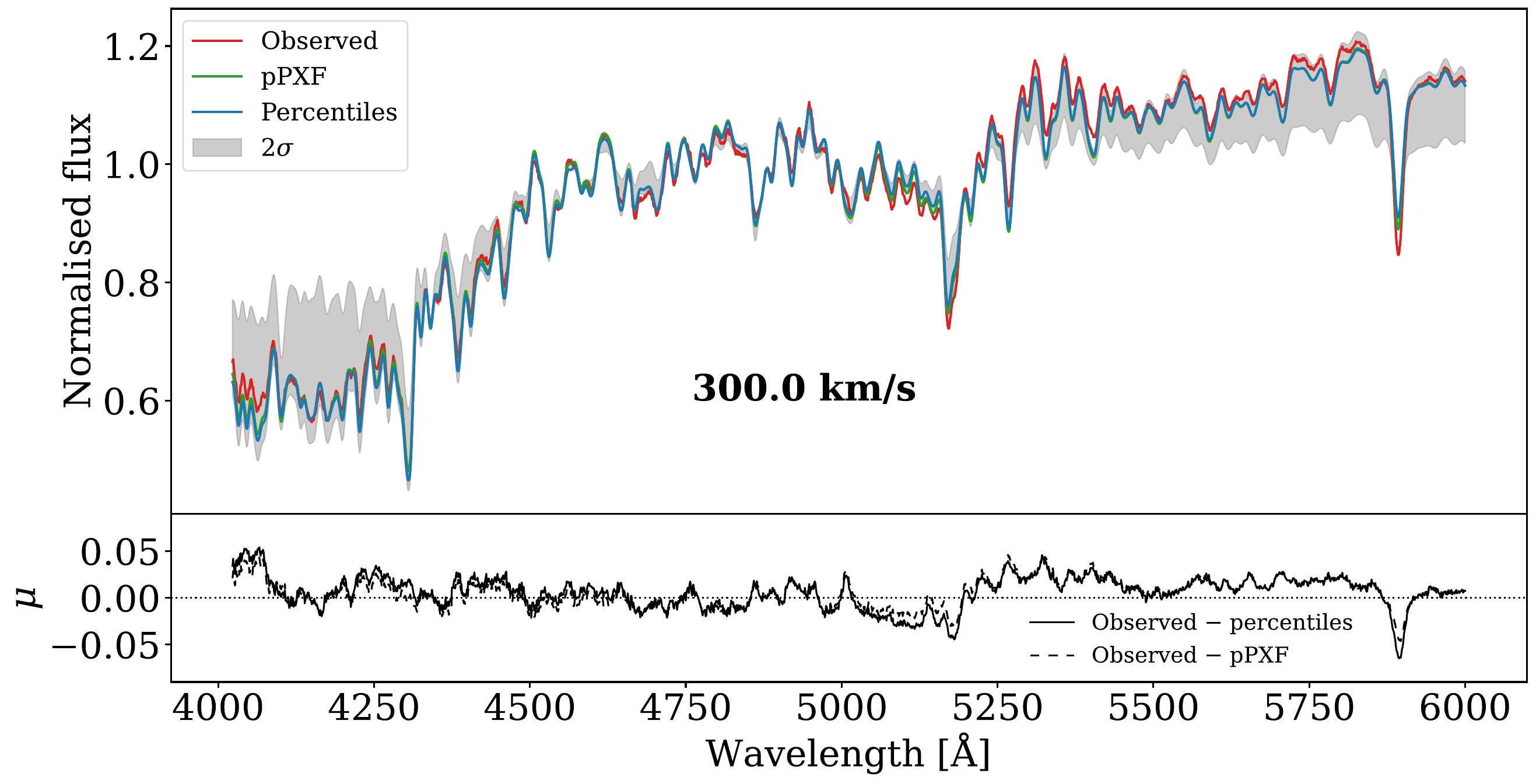}

    \caption{Reconstructions of the observed spectra for the stacks with velocity dispersion $105$, $205,$ and $300$ km/s. In red we show the observed spectra, normalised by the median, in the wavelength range $[4023, 6000] \; \AA$. In blue, we show the spectra obtained as a linear combination of MILES SSP spectra according to our model, using the median values of the posteriors predicted. We repeated the procedure for the median $\pm2\sigma$ to get the grey-shadowed stripe, which shows how the uncertainties in the quantities are manifested in the spectra. In green, we plot the spectra obtained with the weights given by pPXF. In the lower panels, we include the residuals between the observed spectra and those reconstructed from the percentiles and metallicity predicted by our model (solid black line), and between the observed spectra and the one recovered by pPXF (dashed black line). The dotted line indicates the zero level. The typical value of the residuals of our method $(1.7\%)$ is very similar to the average value obtained with pPXF $(1.5\%)$, although in the first case the inference is performed in the latent space, without directly trying to reproduce the observed spectra.}
    \label{spectra_obs}
\end{figure}

We include another reconstruction of the spectra in Fig~\ref{spectra_obs}, namely a linear combination of the same MILES templates of  $34$ different ages and $12$ different metallicities ($408$ templates in total), where the weights are given by the Penalized PiXel-Fitting code \citep[pPXF;][]{ppxf}. The purpose of this comparison is to benchmark our amortised inference against this well-known optimisation method. In order to do that, we used the same wavelength range as before without any continuum correction. The average of the residuals between the observed spectra and the pPXF fitting is $1.5 \%$. The observed spectra, our reconstructions from the percentiles and the reconstructions made with the weights of pPXF are in close agreement, not only in the latent space, but also in the wavelength space. We find that our method generally fits the spectra as accurately as pPXF despite, again, not being designed to reproduce them.\\

 On the other hand, the pPXF fitting takes $\sim30$ s for each stack, and since it does not provide an uncertainty, this measurement would be equivalent to a sample of the posteriors obtained with our model, which take $4 \cdot 10^{-4}$\,s per stack.  Thanks to this acceleration of five orders of magnitude, our model can easily produce $10^3$ or $10^4$ samples of the posteriors for each observation and thus, always under the assumptions of the simulation, obtain a proper calibration of the uncertainty.

%time and deal with uncertainties in pPXF

\begin{table}[h!]
\caption{Inferred properties for 18 stacks of ETGs.}
\centering
\resizebox{\hsize}{!}
            {
\begin{tblr}{
  cells = {c},
  hline{1,20} = {-}{0.08em},
  hline{2} = {-}{0.05em},
}

${\sigma}$ {[km/s]} & {M10\% [Gyr]} & {M50\% [Gyr]} & {M90\% [Gyr]} & $\rm{[M/H]}$ \\
100-110                            & 0.42$\pm0.22$         & 1.21$\pm0.34$        & 2.89$\pm0.63$        & 0.11$\pm0.24$   \\
110-120                            & 0.36$\pm0.22$        & 1.12$\pm0.32$        & 2.63$\pm0.66$        & 0.13$\pm0.23$  \\
120-130                            & 0.32$\pm0.21$        & 0.95$\pm0.33$        & 2.21$\pm0.75$        & -0.05$\pm0.32$ \\
130-140                            & 0.23$\pm0.17$        & 0.72$\pm0.29$        & 1.66$\pm0.70$         & -0.12$\pm0.40$  \\
140-150                            & 0.16$\pm0.13$        & 0.63$\pm0.24$        & 1.43$\pm0.57$        & 0.06$\pm0.38$  \\
150-160                            & 0.17$\pm0.13$        & 0.61$\pm0.25$        & 1.31$\pm0.61$        & -0.05$\pm0.39$ \\
160-170                            & 0.15$\pm0.12$        & 0.59$\pm0.21$        & 1.18$\pm0.54$        & 0.01$\pm0.40$   \\
170-180                            & 0.15$\pm0.11$        & 0.60$\pm0.21$         & 1.11$\pm0.53$        & -0.05$\pm0.40$  \\
180-190                            & 0.13$\pm0.10$         & 0.59$\pm0.20$         & 1.08$\pm0.50$         & -0.07$\pm0.41$ \\
190-200                            & 0.13$\pm0.09$        & 0.62$\pm0.18$        & 1.04$\pm0.46$        & 0.01$\pm0.37$  \\
200-210                            & 0.13$\pm0.11$         & 0.63$\pm0.16$       & 1.02$\pm0.41$        & -0.03$\pm0.39$  \\
210-220                            & 0.13$\pm0.10$         & 0.64$\pm0.17$        & 1.02$\pm0.44$        & -0.04$\pm0.37$ \\
220-230                            & 0.12$\pm0.09$        & 0.65$\pm0.15$        & 1.02$\pm0.42$        & -0.07$\pm0.37$ \\
230-240                            & 0.12$\pm0.12$        & 0.67$\pm0.14$        & 1.06$\pm0.39$        & 0.07$\pm0.37$  \\
240-250                            & 0.12$\pm0.07$        & 0.66$\pm0.12$        & 1.02$\pm0.39$        & 0.03$\pm0.38$  \\
250-260                            & 0.12$\pm0.07$        & 0.69$\pm0.11$        & 0.96$\pm0.36$        & 0.05$\pm0.39$  \\
260-280                            & 0.12$\pm0.06$        & 0.69$\pm0.12$        & 0.99$\pm0.41$        & -0.15$\pm0.38$ \\
280-320                            & 0.12$\pm0.07$         & 0.66$\pm0.11$        & 1.02$\pm0.37$       & 0.11$\pm0.34$

\end{tblr}}
\label{table_obs}

\tablefoot{We include the cosmic time in Gyr at which $10$\%, $50$\%, and $90$\% of the total stellar mass are formed, as well as the metallicity $\rm{[M/H]}$. The values shown correspond to the median values of the predicted posterior distributions, and the uncertainties to the standard deviations.}
\end{table}

%--------------------------------------------------------------------

\section{Discussion}

\label{discussion}

We have developed a new approach to estimating the SFHs of galaxies from their optical absorption spectra, using SBI to obtain posteriors in a fully probabilistic treatment. From interpretable latent representations of the spectra, we inferred the stellar mass  growth and metallicity, reaching high accuracy for the synthetic sample, as well as  properly calibrated uncertainties, evaluated through a SBC test. It requires less than half a second to generate $1{,}000$ samples from the posteriors for each galaxy. This not only makes it possible to address a large dataset, but also to increase the number of samples from the posteriors, raising the precision of the inferred galaxy properties and their uncertainties. \\

\subsection{Model considerations}

Although our SBI framework is substantially faster than other Bayesian inference methods, such as MCMC, and provides well-calibrated uncertainties, unlike classical inversion methods applied to spectral fitting, it is subject to different systematics and modelling assumptions. One of the obvious limitations is the use of a forward model to train the network, because as our understanding of galaxy evolution is incomplete, the considerations taken when generating the synthetic data will never be fully constrained by a typical galaxy spectrum. Therefore, the prior, set by the distribution of parameters in the training set, will always have at least a moderate role in determining the answer.\\

In particular, the simplifications made in terms of chemical evolution, by combining MILES SSP spectra with a fixed metallicity (instead of taking into account the metallicity histories) and working with base $\alpha$-enhancement models, may affect the reliability of our model predictions, given the intrinsic degeneration between the age and the metal content in the spectra. Moreover, there is currently no consensus on the stellar evolution \citep{Conroy_2009} or on the IMF of galaxies \citep{Mart_n_Navarro_2014}, and our model training has been restricted to a specific set of isochrones assuming a particular IMF parametrisation. This limits the range of the training data, and dooms our model to fail in observations of galaxies that differ from these considerations.  Nevertheless, MCMC-based approaches also suffer from these forward model mis-specifications. As shown in \cite{acquaviva11}, both the values of the physical parameters and their uncertainties may depend on the assumptions made in the SPS modelling. The main difference is that as SBI uses a neural network to estimate the posterior, the inference is potentially more sensitive to the discrepancies between the training dataset and the true observations.\\

It is known that the SFHs have a strong impact on the predicted posteriors, as shown by \cite{Leja_2019}, and models that mimic the breadth  distribution of SFR($t$) in the real Universe are required. In this work, in the pipeline provided by the GP-SFH module, we selected a dispersive prior: a Dirichlet distribution with $\alpha=1.0$ for the fractional sSFR in each of the three equally spaced time bins, positioning ourselves in favour of short-term variations in the SFHs (with smaller characteristic times). However, this selection must be explored with caution in future works.

\subsection{Observations}

We inferred the SFH and the metallicity for stacks of SDSS spectra of ETGs. The model recovers the well-known relationship between  age and velocity dispersion, showing that the most massive galaxies ($\sigma \sim 300$ km/s) build up their stellar masses more abruptly, up to 90\% of their total stellar mass by $1$ Gyr after the Big Bang, while the growth of stellar mass is softened in less massive galaxies. How these massive galaxies, up to $M_{*} \sim 10^{12} \, \rm M_{\odot}$ for the highest velocity dispersion stack, can form most of their masses so early is a question that is still open, given that a priori it requires a very high star formation efficiency. These galaxies are rare in the known Universe, with rapid bursts of star formation and followed by rapid quenching. The stars we see may have formed in different progenitor galaxies, which were located in the highest-density peaks of the Cosmic Web, and later assembled in major mergers \citep{conselice2007assembly}. To attend to these processes, hierarchical assembly models are necessary, commonly studied through large-scale cosmological simulations \citep[e.g.][]{Angeloudi_2023}.\\

As future work, we propose studying the SFHs and metallicities for late-type galaxies, for which we expect modelling difficulties due to the larger number of emission lines and potential `over-shining' effects when recovering the metallicity and ages of their first stars to form. So far we have not incorporated emission lines, noise models or dust attenuation, but including these aspects in the future is essential to manage a wide range of observations.  We can model the emission lines with specialised libraries \citep[e.g.][]{cloudy,fado}, or with data-driven approaches, training a neural network to  directly learn the relation between the continuum and the emission features in real observed spectra. This last method can  also be extended to apply realistic noise to the spectra in the simulation, matching the uncertainty distributions in observations, and possibly introducing both the synthetic noisy spectra and the uncertainties for the fluxes into the encoder, resulting in noise-aware latent representations.  In this way, we can condition the Bayesian inference with the noise.\\

Our approach can contribute, given its speed and error handling, to study the biases that produce aspects of the forward model that have traditionally been assumed in similar analyses. We can do that by studying the differences we obtain in the posteriors when changing the specifications and ranges of the parameters in the SPS model. An example would be to change the IMF, which was fixed in this work. By using templates with a bottom-heavy IMF, as expected at least at some stages of the life of massive ETGs \citep{weidner13,demasi,nacho2023}, we obtain a smoother growth of stellar mass for the stacked spectra we analysed, slightly relaxing the requirement of an extremely high star formation efficiency in the first years after the Big Bang.\\

Apart from performing model selection, and following the first steps taken in this project for the SDSS spectra, we plan to expand our model to  perform a deep sampling of current or upcoming spectroscopic surveys such as DESI \citep{desi}, WEAVE \citep{weave}, 4MOST \citep{4most}, or MOONS \citep{MOONS}, which will observe billions of galaxies and produce more than $10^8$ TB of data \citep{2023Smith}. 
%--------------------------------------------------------------------

\section{Summary and outlook}

By analysing the spectrum of a galaxy, one can infer physical properties such as its stellar mass, star formation rates, and chemical abundances, key ingredients for our understanding of galaxy evolution. State-of-the-art spectral fitting methods use MCMC sampling to perform Bayesian statistical inference, deriving the posterior probability distributions of galaxy properties from observations. However, obtaining a posterior with these methods can take $\gtrsim 10-100$ CPU hours per galaxy and thus implies a major bottleneck for addressing large galaxy surveys.\\

We demonstrate in this work that an amortised SBI, with a previous encoding of the spectra, provides an alternative approach for spectral fitting through a neural density estimator. By using MILES templates and non-parametric SFHs, we have constructed a flexible model that recovers SFHs and metallicities with their uncertainties for observed stacks of spectra from  the SDSS, taking $\sim 4$\,s per galaxy to generate $10{,}000$ samples from the posteriors for the ten quantities we estimate. The key results of our project are as follows.\\

First, we constructed a model composed of an encoder for the spectra based on a dot-product attention model, plus a MAF, to estimate the posterior distributions for the cosmic times at which nine stellar mass percentiles are reached, and for the metallicity. We trained it over $\sim4$ hours with $135{,}000$ synthetic samples obtained from SSP MILES templates and non-parametric SFHs from the GP-SFH module.\\
    
 Second, we tested the model with $15{,}000$ synthetic samples, reaching an $R^2$ score of $0.88$, $0.97$, $0.98$, and $0.96$ when estimating  the percentiles $10\%$, $50\%$, and $90\%$ and the metallicity, respectively. The model recovers uncertainties associated with the degeneration of the inversion problem.  It struggles more in assessing the first percentiles in  galaxies with recent star formation but reflects these difficulties adequately in the posteriors, as indicated by an SBC test.\\
 
Third, we estimated with our model the SFHs and metallicities of $18$ stacks of ETGs from the SDSS with velocity dispersions in the range $105-300$ km/s, as well as  reliable uncertainties for the inferred properties. We uncover very early bursts of star formation in the most massive galaxies and a smoother growth of stellar mass when moving to intermediate masses, recovering the well-known relation between age and velocity dispersion. Moreover, we accurately reconstructed the spectra from MILES templates using the nine ages and the mean metallicity measured, in good agreement with the real spectra, with an averaged error of $1.7 \%$.\\

In the future, we plan to explore different sets of assumptions for the isochrones, the stellar library, and the IMF, as well as the effects of the priors introduced by the selection of SFHs and metallicity ranges. To simulate the spectra of young stellar populations, a data-driven approach, possibly a neural network, can be used to directly learn the emission features from observed spectra. This methodology can also be extended to model the noise. To obtain noise-aware representations, the encoder can use both the uncertainties in the fluxes and the noisy spectra. At that stage, the optimal dimension of the latent vectors must be studied again. Such a complete model would allow one to analyse massive sets of galaxy spectra in a highly efficient and reliable manner.\\

The entire pipeline, including the scripts for the simulation and the Bayesian inference framework, is publicly available at \href{https://github.com/patriglesias/SBI_SFHs.git}{GitHub.}\footnote{\href{https://github.com/patriglesias/SBI_SFHs.git}{https://github.com/patriglesias/SBI\_SFHs.git}}

%--------------------------------------------------------------------

\begin{acknowledgements}

Co-funded by the European Union. Views and opinions expressed are however those of the author(s) only and do not necessarily reflect those of the European Union. Neither the European Union nor the granting authority can be held responsible for them. MHC and PIN acknowledge financial support from the State Research Agency of the Spanish Ministry of Science and Innovation (AEI-MCINN) under the grants ``Galaxy Evolution with Artificial Intelligence'' with reference PGC2018-100852-A-I00 and ``BASALT'' with reference PID2021-126838NB-I00. IMN acknowledges support from AEI-MCINN under the grant ``B-CycleS'' with reference PID2022-140869NB-I00. JHK acknowledges support from AEI-MCINN under the grant ``The structure and evolution of galaxies and their outer regions'' and the European Regional Development Fund (ERDF) with reference PID2022-136505NB-I00/10.13039/501100011033. EP acknowledges support from the Turing Scheme.

\end{acknowledgements}

\bibliographystyle{aa}
\bibliography{references}

\begin{thebibliography}{73}
\expandafter\ifx\csname natexlab\endcsname\relax\def\natexlab#1{#1}\fi

\bibitem[{{Acquaviva} {et~al.}(2011){Acquaviva}, {Gawiser}, \& {Guaita}}]{acquaviva11}
{Acquaviva}, V., {Gawiser}, E., \& {Guaita}, L. 2011, \apj, 737, 47

\bibitem[{Ade {et~al.}(2014)Ade, Aghanim, Armitage-Caplan, Arnaud, Ashdown, Atrio-Barandela, Aumont, Baccigalupi, Banday, Barreiro, Bartlett, Battaner, Benabed, Beno{\^{\i} }t, Benoit-L{\'{e}}vy, Bernard, Bersanelli, Bielewicz, Bobin, Bock, Bonaldi, Bond, Borrill, Bouchet, Bridges, Bucher, Burigana, Butler, Calabrese, Cappellini, Cardoso, Catalano, Challinor, Chamballu, Chary, Chen, Chiang, Chiang, Christensen, Church, Clements, Colombi, Colombo, Couchot, Coulais, Crill, Curto, Cuttaia, Danese, Davies, Davis, de~Bernardis, de~Rosa, de~Zotti, Delabrouille, Delouis, D{\'{e}}sert, Dickinson, Diego, Dolag, Dole, Donzelli, Dor{\'{e}}, Douspis, Dunkley, Dupac, Efstathiou, Elsner, En{\ss}lin, Eriksen, Finelli, Forni, Frailis, Fraisse, Franceschi, Gaier, Galeotta, Galli, Ganga, Giard, Giardino, Giraud-H{\'{e}}raud, Gjerl{\o}w, Gonz{\'{a}}lez-Nuevo, G{\'{o}}rski, Gratton, Gregorio, Gruppuso, Gudmundsson, Haissinski, Hamann, Hansen, Hanson, Harrison, Henrot-Versill{\'{e}}, Hern{\'{a}}ndez-Monteagudo, Herranz, Hildebrandt, Hivon, Hobson, Holmes, Hornstrup, Hou, Hovest, Huffenberger, Jaffe, Jaffe, Jewell, Jones, Juvela, Keihänen, Keskitalo, Kisner, Kneissl, Knoche, Knox, Kunz, Kurki-Suonio, Lagache, Lähteenmäki, Lamarre, Lasenby, Lattanzi, Laureijs, Lawrence, Leach, Leahy, Leonardi, Le{\'{o}}n-Tavares, Lesgourgues, Lewis, Liguori, Lilje, Linden-V{\o}rnle, L{\'{o}}pez-Caniego, Lubin, Mac{\'{\i}}as-P{\'{e}}rez, Maffei, Maino, Mandolesi, Maris, Marshall, Martin, Mart{\'{\i}}nez-Gonz{\'{a}}lez, Masi, Massardi, Matarrese, Matthai, Mazzotta, Meinhold, Melchiorri, Melin, Mendes, Menegoni, Mennella, Migliaccio, Millea, Mitra, Miville-Desch{\^{e}}nes, Moneti, Montier, Morgante, Mortlock, Moss, Munshi, Murphy, Naselsky, Nati, Natoli, Netterfield, N{\o}rgaard-Nielsen, Noviello, Novikov, Novikov, O'Dwyer, Osborne, Oxborrow, Paci, Pagano, Pajot, Paladini, Paoletti, Partridge, Pasian, Patanchon, Pearson, Pearson, Peiris, Perdereau, Perotto, Perrotta, Pettorino, Piacentini, Piat, Pierpaoli, Pietrobon, Plaszczynski, Platania, Pointecouteau, Polenta, Ponthieu, Popa, Poutanen, Pratt, Pr{\'{e}}zeau, Prunet, Puget, Rachen, Reach, Rebolo, Reinecke, Remazeilles, Renault, Ricciardi, Riller, Ristorcelli, Rocha, Rosset, Roudier, Rowan-Robinson, Rubi{\~{n}}o-Mart{\'{\i}}n, Rusholme, Sandri, Santos, Savelainen, Savini, Scott, Seiffert, Shellard, Spencer, Starck, Stolyarov, Stompor, Sudiwala, Sunyaev, Sureau, Sutton, Suur-Uski, Sygnet, Tauber, Tavagnacco, Terenzi, Toffolatti, Tomasi, Tristram, Tucci, Tuovinen, Türler, Umana, Valenziano, Valiviita, Tent, Vielva, Villa, Vittorio, Wade, Wandelt, Wehus, White, White, Wilkinson, Yvon, Zacchei, \& Zonca}]{planck13}
Ade, P. A.~R., Aghanim, N., Armitage-Caplan, C., {et~al.} 2014, \aap, 571, A16

\bibitem[{Angeloudi {et~al.}(2023)Angeloudi, Falcón-Barroso, Huertas-Company, Sarmiento, Pillepich, Walo-Mart{\'{\i}}n, \& Eisert}]{Angeloudi_2023}
Angeloudi, E., Falcón-Barroso, J., Huertas-Company, M., {et~al.} 2023, \mnras

\bibitem[{Cameron \& Pettitt(2012)}]{Cameron_2012}
Cameron, E. \& Pettitt, A.~N. 2012, \mnras, 425, 44

\bibitem[{{Cappellari}(2023)}]{Cappellari2022}
{Cappellari}, M. 2023, \mnras, 526, 3273

\bibitem[{{Cappellari} \& {Emsellem}(2004)}]{ppxf}
{Cappellari}, M. \& {Emsellem}, E. 2004, \pasp, 116, 138

\bibitem[{{Cardoso} {et~al.}(2019){Cardoso}, {Gomes}, \& {Papaderos}}]{fado}
{Cardoso}, L. S.~M., {Gomes}, J.~M., \& {Papaderos}, P. 2019, \aap, 622, A56

\bibitem[{Carnall {et~al.}(2019)Carnall, Leja, Johnson, McLure, Dunlop, \& Conroy}]{Carnall_2019}
Carnall, A.~C., Leja, J., Johnson, B.~D., {et~al.} 2019, \apj, 873, 44

\bibitem[{Cenarro {et~al.}(2001)Cenarro, Cardiel, Gorgas, Peletier, Vazdekis, \& Prada}]{Cenarro_2001}
Cenarro, A.~J., Cardiel, N., Gorgas, J., {et~al.} 2001, \mnras, 326, 959

\bibitem[{Cervantes \& Vazdekis(2008)}]{cervantes09}
Cervantes, J.~L. \& Vazdekis, A. 2008, \mnras, 392, 691

\bibitem[{Chabrier(2003)}]{Chabrier_2003}
Chabrier, G. 2003, \pasp, 115, 763

\bibitem[{Choi {et~al.}(2016)Choi, Dotter, Conroy, Cantiello, Paxton, \& Johnson}]{Choi_2016}
Choi, J., Dotter, A., Conroy, C., {et~al.} 2016, \apj, 823, 102

\bibitem[{Cirasuolo {et~al.}(2020)Cirasuolo, Fairley, Rees, Gonzalez, Taylor, Maiolino, Afonso, Evans, Flores, Lilly, Oliva, Paltani, Vanzi, Abreu, Accardo, Adams, Álvarez Méndez, Amans, Amarantidis, Atek, Atkinson, Banerji, Barrett, Barrientos, Bauer, Beard, Béchet, Belfiore, Bellazzini, Benoist, Best, Biazzo, Black, Boettger, Bonifacio, Bowler, Bragaglia, Brierley, Brinchmann, Brinkmann, Buat, Buitrago, Burgarella, Burningham, Buscher, Cabral, Caffau, Cardoso, Carnall, Carollo, Castillo, Castignani, Catelan, Cicone, Cimatti, Cioni, Clementini, Cochrane, Coelho, Colling, Contini, Contreras, Conzelmann, Cresci, Cropper, Cucciati, Cullen, Cumani, Curti, Da~Silva, Daddi, Dalessandro, Dalessio, Dauvin, Davidson, De~Laverny, Delplancke-Ströbele, De~Lucia, Del~Vecchio, Dessauges-Zavadsky, Di~Matteo, Dole, Drass, Dunlop, Dünner, Eales, Ellis, Enriques, Fasola, Ferguson, Ferruzzi, Fisher, Flores, Fontana, Forchi, Francois, Franzetti, Gargiulo, Garilli, Gaudemard, Gieles, Gilmore, Ginolfi, Gomes, Guinouard, Gutierrez, Haigron, Hammer, Hammersley, Haniff, Harrison, Haywood, Hill, Hubin, Humphrey, Ibata, Infante, Ives, Ivison, Iwert, Jablonka, Jakob, Jarvis, King, Kneib, Laporte, Lawrence, Lee, Li~Causi, Lorenzoni, Lucatello, Luco, Macleod, Magliocchetti, Magrini, Mainieri, Maire, Mannucci, Martin, Matute, Maurogordato, McGee, Mcleod, McLure, McMahon, Melse, Messias, Mucciarelli, Nisini, Nix, Norberg, Oesch, Oliveira, Origlia, Padilla, Palsa, Pancino, Papaderos, Pappalardo, Parry, Pasquini, Peacock, Pedichini, Pello, Peng, Pentericci, Pfuhl, Piazzesi, Popovic, Pozzetti, Puech, Puzia, Raichoor, Randich, Recio-Blanco, Reis, Reix, Renzini, Rodrigues, Rojas, Rojas-Arriagada, Rota, Royer, Sacco, Sanchez-Janssen, Sanna, Santos, Sarzi, Schaerer, Schiavon, Schnell, Schultheis, Scodeggio, Serjeant, Shen, Simmonds, Smoker, Sobral, Sordet, Spérone, Strachan, Sun, Swinbank, Tait, Tereno, Tojeiro, Torres, Tosi, Tozzi, Tresiter, Valenti, Valenzuela~Navarro, Vanzella, Vergani, Verhamme, Vernet, Vignali, Vinther, Von~Dran, Waring, Watson, Wild, Willesme, Woodward, Wuyts, Yang, Zamorani, Zoccali, Bluck, \& Trussler}]{MOONS}
Cirasuolo, M., Fairley, A., Rees, P., {et~al.} 2020, The Messenger, pp. 10-17, June 2020.

\bibitem[{Conroy(2013)}]{Conroy_2013}
Conroy, C. 2013, \araa, 51, 393

\bibitem[{Conroy {et~al.}(2009)Conroy, Gunn, \& White}]{Conroy_2009}
Conroy, C., Gunn, J.~E., \& White, M. 2009, \apj, 699, 486

\bibitem[{{Conselice}(2008)}]{conselice2007assembly}
{Conselice}, C.~J. 2008, \aspc, 390, 403

\bibitem[{{Dalton} {et~al.}(2012){Dalton}, {Trager}, {Abrams}, {Carter}, {Bonifacio}, {Aguerri}, {MacIntosh}, {Evans}, {Lewis}, {Navarro}, {Agocs}, {Dee}, {Rousset}, {Tosh}, {Middleton}, {Pragt}, {Terrett}, {Brock}, {Benn}, {Verheijen}, {Cano Infantes}, {Bevil}, {Steele}, {Mottram}, {Bates}, {Gribbin}, {Rey}, {Rodriguez}, {Delgado}, {Guinouard}, {Walton}, {Irwin}, {Jagourel}, {Stuik}, {Gerlofsma}, {Roelfsma}, {Skillen}, {Ridings}, {Balcells}, {Daban}, {Gouvret}, {Venema}, \& {Girard}}]{weave}
{Dalton}, G., {Trager}, S.~C., {Abrams}, D.~C., {et~al.} 2012, in Society of Photo-Optical Instrumentation Engineers (SPIE) Conference Series, Vol. 8446, Ground-based and Airborne Instrumentation for Astronomy IV, ed. I.~S. {McLean}, S.~K. {Ramsay}, \& H.~{Takami}, 84460P

\bibitem[{{de Jong}(2011)}]{4most}
{de Jong}, R. 2011, The Messenger, 145, 14

\bibitem[{{De Masi} {et~al.}(2019){De Masi}, {Vincenzo}, {Matteucci}, {Rosani}, {La Barbera}, {Pasquali}, \& {Spitoni}}]{demasi}
{De Masi}, C., {Vincenzo}, F., {Matteucci}, F., {et~al.} 2019, \mnras, 483, 2217

\bibitem[{{Falc{\'o}n-Barroso} {et~al.}(2011){Falc{\'o}n-Barroso}, {S{\'a}nchez-Bl{\'a}zquez}, {Vazdekis}, {Ricciardelli}, {Cardiel}, {Cenarro}, {Gorgas}, \& {Peletier}}]{falcon11}
{Falc{\'o}n-Barroso}, J., {S{\'a}nchez-Bl{\'a}zquez}, P., {Vazdekis}, A., {et~al.} 2011, \aap, 532, A95

\bibitem[{{Ferland} {et~al.}(1998){Ferland}, {Korista}, {Verner}, {Ferguson}, {Kingdon}, \& {Verner}}]{cloudy}
{Ferland}, G.~J., {Korista}, K.~T., {Verner}, D.~A., {et~al.} 1998, \pasp, 110, 761

\bibitem[{Fernandes {et~al.}(2005)Fernandes, Mateus, Sodré, Stasińska, \& Gomes}]{starlight}
Fernandes, R.~C., Mateus, A., Sodré, L., Stasińska, G., \& Gomes, J.~M. 2005, \mnras, 358, 363

\bibitem[{{Franx} \& {Illingworth}(1990)}]{Franx90}
{Franx}, M. \& {Illingworth}, G. 1990, \apjl, 359, L41

\bibitem[{Girardi {et~al.}(2000)Girardi, Bressan, Bertelli, \& Chiosi}]{Girardi_2000}
Girardi, L., Bressan, A., Bertelli, G., \& Chiosi, C. 2000, \aaps, 141, 371

\bibitem[{Gonneau {et~al.}(2020)Gonneau, Lyubenova, Lancon, Trager, Peletier, Arentsen, Chen, Coelho, Dries, Falc{\'{o}}n-Barroso, Prugniel, S{\'{a}}nchez-Bl{\'{a}}zquez, Vazdekis, \& Verro}]{Gonneau_2020}
Gonneau, A., Lyubenova, M., Lancon, A., {et~al.} 2020, \aap, 634, A133

\bibitem[{{Hahn} {et~al.}(2023{\natexlab{a}}){Hahn}, {Eickenberg}, {Ho}, {Hou}, {Lemos}, {Massara}, {Modi}, {Moradinezhad Dizgah}, {R{\'e}galdo-Saint Blancard}, \& {Abidi}}]{hahn23}
{Hahn}, C., {Eickenberg}, M., {Ho}, S., {et~al.} 2023{\natexlab{a}}, \jcap, 2023, 010

\bibitem[{{Hahn} {et~al.}(2023{\natexlab{b}}){Hahn}, {Kwon}, {Tojeiro}, {Siudek}, {Canning}, {Mezcua}, {Tinker}, {Brooks}, {Doel}, {Fanning}, {Gazta{\~n}aga}, {Kehoe}, {Landriau}, {Meisner}, {Moustakas}, {Poppett}, {Tarle}, {Weiner}, \& {Zou}}]{desi}
{Hahn}, C., {Kwon}, K.~J., {Tojeiro}, R., {et~al.} 2023{\natexlab{b}}, \apj, 945, 16

\bibitem[{Hahn \& Melchior(2022)}]{Hahn_2022}
Hahn, C. \& Melchior, P. 2022, \apj, 938, 11

\bibitem[{Hinton \& Salakhutdinov(2006)}]{hinton}
Hinton, G.~E. \& Salakhutdinov, R.~R. 2006, Science, 313, 504

\bibitem[{{Huertas-Company} {et~al.}(2023){Huertas-Company}, {Sarmiento}, \& {Knapen}}]{huertascompany2023brief}
{Huertas-Company}, M., {Sarmiento}, R., \& {Knapen}, J.~H. 2023, RAS Techniques and Instruments, 2, 441

\bibitem[{Hunt {et~al.}(2024)Hunt, Pimbblet, \& Benoit}]{hunt2024predicting}
Hunt, L.~J., Pimbblet, K.~A., \& Benoit, D.~M. 2024, \mnras, stae479

\bibitem[{Iyer \& Gawiser(2017)}]{Iyer_2017}
Iyer, K. \& Gawiser, E. 2017, \apj, 838, 127

\bibitem[{Johnson {et~al.}(2021)Johnson, Leja, Conroy, \& Speagle}]{Johnson_2021}
Johnson, B.~D., Leja, J., Conroy, C., \& Speagle, J.~S. 2021, \apjs, 254, 22

\bibitem[{{Khullar} {et~al.}(2022){Khullar}, {Nord}, {{\'C}iprijanovi{\'c}}, {Poh}, \& {Xu}}]{gourav22}
{Khullar}, G., {Nord}, B., {{\'C}iprijanovi{\'c}}, A., {Poh}, J., \& {Xu}, F. 2022, Machine Learning: Science and Technology, 3, 04LT04

\bibitem[{{Kramer}(1991)}]{kramer91}
{Kramer}, M.~A. 1991, AIChE Journal, 37, 233

\bibitem[{Kroupa(2001)}]{Kroupa_2001}
Kroupa, P. 2001, \mnras, 322, 231

\bibitem[{{Kwon} \& {Hahn}(2024)}]{kwon24}
{Kwon}, K.~J. \& {Hahn}, C. 2024, submitted to ApJ, arXiv:2401.12318

\bibitem[{{La Barbera} {et~al.}(2013){La Barbera}, Ferreras, Vazdekis, de~la Rosa, de~Carvalho, Trevisan, Falc{\'{o} }n-Barroso, \& Ricciardelli}]{La_Barbera_2013}
{La Barbera}, F., Ferreras, I., Vazdekis, A., {et~al.} 2013, \mnras, 433, 3017

\bibitem[{Leja {et~al.}(2019)Leja, Carnall, Johnson, Conroy, \& Speagle}]{Leja_2019}
Leja, J., Carnall, A.~C., Johnson, B.~D., Conroy, C., \& Speagle, J.~S. 2019, \apj, 876, 3

\bibitem[{{Lilly} {et~al.}(2013){Lilly}, {Carollo}, {Pipino}, {Renzini}, \& {Peng}}]{lilly13}
{Lilly}, S.~J., {Carollo}, C.~M., {Pipino}, A., {Renzini}, A., \& {Peng}, Y. 2013, \apj, 772, 119

\bibitem[{Lovell {et~al.}(2019)Lovell, Acquaviva, Thomas, Iyer, Gawiser, \& Wilkins}]{lovell19}
Lovell, C.~C., Acquaviva, V., Thomas, P.~A., {et~al.} 2019, \mnras, 490, 5503

\bibitem[{{Maksymowicz-Maciata} {et~al.}(2024){Maksymowicz-Maciata}, {Spiniello}, {Mart{\'\i}n-Navarro}, {Ferr{\'e}-Mateu}, {Bevacqua}, {Cappellari}, {D'Ago}, {Tortora}, {Arnaboldi}, {Hartke}, {Napolitano}, {Saracco}, \& {Scognamiglio}}]{2024maciata}
{Maksymowicz-Maciata}, M., {Spiniello}, C., {Mart{\'\i}n-Navarro}, I., {et~al.} 2024, \mnras, 531, 2864

\bibitem[{{Mart{\'\i}n-Navarro} {et~al.}(2023){Mart{\'\i}n-Navarro}, {Spiniello}, {Tortora}, {Coccato}, {D'Ago}, {Ferr{\'e}-Mateu}, {Pulsoni}, {Hartke}, {Arnaboldi}, {Hunt}, {Napolitano}, {Scognamiglio}, \& {Spavone}}]{nacho2023}
{Mart{\'\i}n-Navarro}, I., {Spiniello}, C., {Tortora}, C., {et~al.} 2023, \mnras, 521, 1408

\bibitem[{{Martín-Navarro} {et~al.}(2020){Martín-Navarro}, {Burchett}, \& {Mezcua}}]{nacho}
{Martín-Navarro}, I., {Burchett}, J.~N., \& {Mezcua}, M. 2020, \mnras, 491, 1311

\bibitem[{Martín-Navarro {et~al.}(2015)Martín-Navarro, {La Barbera}, Vazdekis, Falcón-Barroso, \& Ferreras}]{Mart_n_Navarro_2014}
Martín-Navarro, I., {La Barbera}, F., Vazdekis, A., Falcón-Barroso, J., \& Ferreras, I. 2015, \mnras, 447, 1033

\bibitem[{Martín-Navarro {et~al.}(2019)Martín-Navarro, van de Ven, \& Yıldırım}]{Mart_n_Navarro_2019}
Martín-Navarro, I., van de Ven, G., \& Yıldırım, A. 2019, \mnras, 487, 4939–4950

\bibitem[{McInnes {et~al.}(2018)McInnes, Healy, Saul, \& Großberger}]{umap}
McInnes, L., Healy, J., Saul, N., \& Großberger, L. 2018, JOSS, 3, 861

\bibitem[{{Melchior} {et~al.}(2023){Melchior}, {Liang}, {Hahn}, \& {Goulding}}]{melchior2022}
{Melchior}, P., {Liang}, Y., {Hahn}, C., \& {Goulding}, A. 2023, \aj, 166, 74

\bibitem[{Mishra-Sharma(2022)}]{mishrasharma2022inferring}
Mishra-Sharma, S. 2022, Mach. learn. : sci. technol., 3, 01LT03

\bibitem[{Moser {et~al.}(2024)Moser, Kacprzak, Fischbacher, Refregier, Grimm, \& Tortorelli}]{moser24}
Moser, B., Kacprzak, T., Fischbacher, S., {et~al.} 2024, JCAP, 05, 049

\bibitem[{{Ocvirk} {et~al.}(2006){Ocvirk}, {Pichon}, {Lan{\c{c}}on}, \& {Thi{\'e}baut}}]{Ocvirk}
{Ocvirk}, P., {Pichon}, C., {Lan{\c{c}}on}, A., \& {Thi{\'e}baut}, E. 2006, \mnras, 365, 74

\bibitem[{Papamakarios {et~al.}(2017)Papamakarios, Pavlakou, \& Murray}]{papamakarios2018masked}
Papamakarios, G., Pavlakou, T., \& Murray, I. 2017, arXiv e-prints, arXiv:1705.07057

\bibitem[{Pietrinferni {et~al.}(2006)Pietrinferni, Cassisi, Salaris, \& Castelli}]{Pietrinferni_2006}
Pietrinferni, A., Cassisi, S., Salaris, M., \& Castelli, F. 2006, \apj, 642, 797

\bibitem[{Portillo {et~al.}(2020)Portillo, Parejko, Vergara, \& Connolly}]{Portillo_2020}
Portillo, S. K.~N., Parejko, J.~K., Vergara, J.~R., \& Connolly, A.~J. 2020, \apj, 160, 45

\bibitem[{{Saleh} \& {Ehsanes Saleh}(2022)}]{logcosh}
{Saleh}, R.~A. \& {Ehsanes Saleh}, A.~K.~M. 2022, arXiv e-prints, arXiv:2208.04564

\bibitem[{{Salpeter}(1955)}]{salpeter95}
{Salpeter}, E.~E. 1955, \apj, 121, 161

\bibitem[{{Scoville} {et~al.}(2007){Scoville}, {Aussel}, {Brusa}, {Capak}, {Carollo}, {Elvis}, {Giavalisco}, {Guzzo}, {Hasinger}, {Impey}, {Kneib}, {LeFevre}, {Lilly}, {Mobasher}, {Renzini}, {Rich}, {Sanders}, {Schinnerer}, {Schminovich}, {Shopbell}, {Taniguchi}, \& {Tyson}}]{cosmos}
{Scoville}, N., {Aussel}, H., {Brusa}, M., {et~al.} 2007, \apjs, 172, 1

\bibitem[{{Silk} \& {Mamon}(2012)}]{Silk_2012}
{Silk}, J. \& {Mamon}, G.~A. 2012, \raa, 12, 917

\bibitem[{{Smith} \& {Geach}(2023)}]{2023Smith}
{Smith}, M.~J. \& {Geach}, J.~E. 2023, R. Soc. Open Sci., 10, 221454

\bibitem[{{Tacchella} {et~al.}(2022){Tacchella}, {Conroy}, {Faber}, {Johnson}, {Leja}, {Barro}, {Cunningham}, {Deason}, {Guhathakurta}, {Guo}, {Hernquist}, {Koo}, {McKinnon}, {Rockosi}, {Speagle}, {van Dokkum}, \& {Yesuf}}]{tacchela21}
{Tacchella}, S., {Conroy}, C., {Faber}, S.~M., {et~al.} 2022, \apj, 926, 134

\bibitem[{Talts {et~al.}(2018)Talts, Betancourt, Simpson, Vehtari, \& Gelman}]{talts2020validating}
Talts, S., Betancourt, M., Simpson, D., Vehtari, A., \& Gelman, A. 2018, arXiv e-prints, arXiv:1804.06788

\bibitem[{Teimoorinia {et~al.}(2022)Teimoorinia, Archinuk, Woo, Shishehchi, \& Bluck}]{Teimoorinia_2022}
Teimoorinia, H., Archinuk, F., Woo, J., Shishehchi, S., \& Bluck, A. F.~L. 2022, \apj, 163, 71

\bibitem[{Tejero-Cantero {et~al.}(2020)Tejero-Cantero, Boelts, Deistler, Lueckmann, Durkan, Gonçalves, Greenberg, \& Macke}]{tejerocantero2020sbi}
Tejero-Cantero, A., Boelts, J., Deistler, M., {et~al.} 2020, JOSS, 5, 2505

\bibitem[{{Thomas} {et~al.}(2003){Thomas}, {Maraston}, \& {Bender}}]{thomas03}
{Thomas}, D., {Maraston}, C., \& {Bender}, R. 2003, \mnras, 339, 897

\bibitem[{Tojeiro {et~al.}(2007)Tojeiro, Heavens, Jimenez, \& Panter}]{vespa}
Tojeiro, R., Heavens, A.~F., Jimenez, R., \& Panter, B. 2007, \mnras, 381, 1252

\bibitem[{Vazdekis {et~al.}(2010)Vazdekis, Sánchez-Bl{\'{a}}zquez, Falc{\'{o}}n-Barroso, Cenarro, Beasley, Cardiel, Gorgas, \& Peletier}]{Vazdekis_2010}
Vazdekis, A., Sánchez-Bl{\'{a}}zquez, P., Falc{\'{o}}n-Barroso, J., {et~al.} 2010, \mnras

\bibitem[{{Wang} {et~al.}(2024){Wang}, {Leja}, {Labb{\'e}}, {Bezanson}, {Whitaker}, {Brammer}, {Furtak}, {Weaver}, {Price}, {Zitrin}, {Atek}, {Coe}, {Cutler}, {Dayal}, {van Dokkum}, {Feldmann}, {Marchesini}, {Franx}, {F{\"o}rster Schreiber}, {Fujimoto}, {Geha}, {Glazebrook}, {de Graaff}, {Greene}, {Juneau}, {Kassin}, {Kriek}, {Khullar}, {Maseda}, {Mowla}, {Muzzin}, {Nanayakkara}, {Nelson}, {Oesch}, {Pacifici}, {Pan}, {Papovich}, {Setton}, {Shapley}, {Smit}, {Stefanon}, {Suess}, {Taylor}, \& {Williams}}]{2024wang}
{Wang}, B., {Leja}, J., {Labb{\'e}}, I., {et~al.} 2024, \apjs, 270, 12

\bibitem[{{Weidner} {et~al.}(2013){Weidner}, {Ferreras}, {Vazdekis}, \& {La Barbera}}]{weidner13}
{Weidner}, C., {Ferreras}, I., {Vazdekis}, A., \& {La Barbera}, F. 2013, \mnras, 435, 2274

\bibitem[{{Wilkinson} {et~al.}(2017){Wilkinson}, {Maraston}, {Goddard}, {Thomas}, \& {Parikh}}]{firefly}
{Wilkinson}, D.~M., {Maraston}, C., {Goddard}, D., {Thomas}, D., \& {Parikh}, T. 2017, \mnras, 472, 4297

\bibitem[{{Woo} {et~al.}(2024){Woo}, {Walters}, {Archinuk}, {Faber}, {Ellison}, {Teimoorinia}, \& {Iyer}}]{joanna24}
{Woo}, J., {Walters}, D., {Archinuk}, F., {et~al.} 2024, \mnras, 530, 4260

\bibitem[{{Worthey} {et~al.}(1992){Worthey}, {Faber}, \& {Gonzalez}}]{worthey92}
{Worthey}, G., {Faber}, S.~M., \& {Gonzalez}, J.~J. 1992, \apj, 398, 69

\bibitem[{{York} {et~al.}(2000){York}, {Adelman}, {Anderson}, {Anderson}, {Annis}, {Bahcall}, {Bakken}, {Barkhouser}, {Bastian}, {Berman}, {Boroski}, {Bracker}, {Briegel}, {Briggs}, {Brinkmann}, {Brunner}, {Burles}, {Carey}, {Carr}, {Castander}, {Chen}, {Colestock}, {Connolly}, {Crocker}, {Csabai}, {Czarapata}, {Davis}, {Doi}, {Dombeck}, {Eisenstein}, {Ellman}, {Elms}, {Evans}, {Fan}, {Federwitz}, {Fiscelli}, {Friedman}, {Frieman}, {Fukugita}, {Gillespie}, {Gunn}, {Gurbani}, {de Haas}, {Haldeman}, {Harris}, {Hayes}, {Heckman}, {Hennessy}, {Hindsley}, {Holm}, {Holmgren}, {Huang}, {Hull}, {Husby}, {Ichikawa}, {Ichikawa}, {Ivezi{\'c}}, {Kent}, {Kim}, {Kinney}, {Klaene}, {Kleinman}, {Kleinman}, {Knapp}, {Korienek}, {Kron}, {Kunszt}, {Lamb}, {Lee}, {Leger}, {Limmongkol}, {Lindenmeyer}, {Long}, {Loomis}, {Loveday}, {Lucinio}, {Lupton}, {MacKinnon}, {Mannery}, {Mantsch}, {Margon}, {McGehee}, {McKay}, {Meiksin}, {Merelli}, {Monet}, {Munn}, {Narayanan}, {Nash}, {Neilsen}, {Neswold}, {Newberg}, {Nichol}, {Nicinski}, {Nonino}, {Okada}, {Okamura}, {Ostriker}, {Owen}, {Pauls}, {Peoples}, {Peterson}, {Petravick}, {Pier}, {Pope}, {Pordes}, {Prosapio}, {Rechenmacher}, {Quinn}, {Richards}, {Richmond}, {Rivetta}, {Rockosi}, {Ruthmansdorfer}, {Sandford}, {Schlegel}, {Schneider}, {Sekiguchi}, {Sergey}, {Shimasaku}, {Siegmund}, {Smee}, {Smith}, {Snedden}, {Stone}, {Stoughton}, {Strauss}, {Stubbs}, {SubbaRao}, {Szalay}, {Szapudi}, {Szokoly}, {Thakar}, {Tremonti}, {Tucker}, {Uomoto}, {Vanden Berk}, {Vogeley}, {Waddell}, {Wang}, {Watanabe}, {Weinberg}, {Yanny}, {Yasuda}, \& {SDSS Collaboration}}]{SDSS}
{York}, D.~G., {Adelman}, J., {Anderson}, John~E., J., {et~al.} 2000, \aj, 120, 1579

\bibitem[{{Zhang} {et~al.}(2023){Zhang}, {Jayasinghe}, \& {Bloom}}]{zhang23}
{Zhang}, K., {Jayasinghe}, T., \& {Bloom}, J. 2023, in Machine Learning for Astrophysics, 39

\end{thebibliography}
% WARNING
%-------------------------------------------------------------------
% Please note that we have included the references to the file aa.dem in
% order to compile it, but we ask you to:
%
% - use BibTeX with the regular commands:
%   \bibliographystyle{aa} % style aa.bst
%   \bibliography{Yourfile} % your references Yourfile.bib
%
% - join the .bib files when you upload your source files
%-------------------------------------------------------------------

\begin{appendix}

\section{Low-dimensional representations of the spectra}
\label{encoding}

\begin{figure}[h]
\begin{minipage}[t]{\textwidth}
    \centering
    \includegraphics[width=0.33\textwidth]{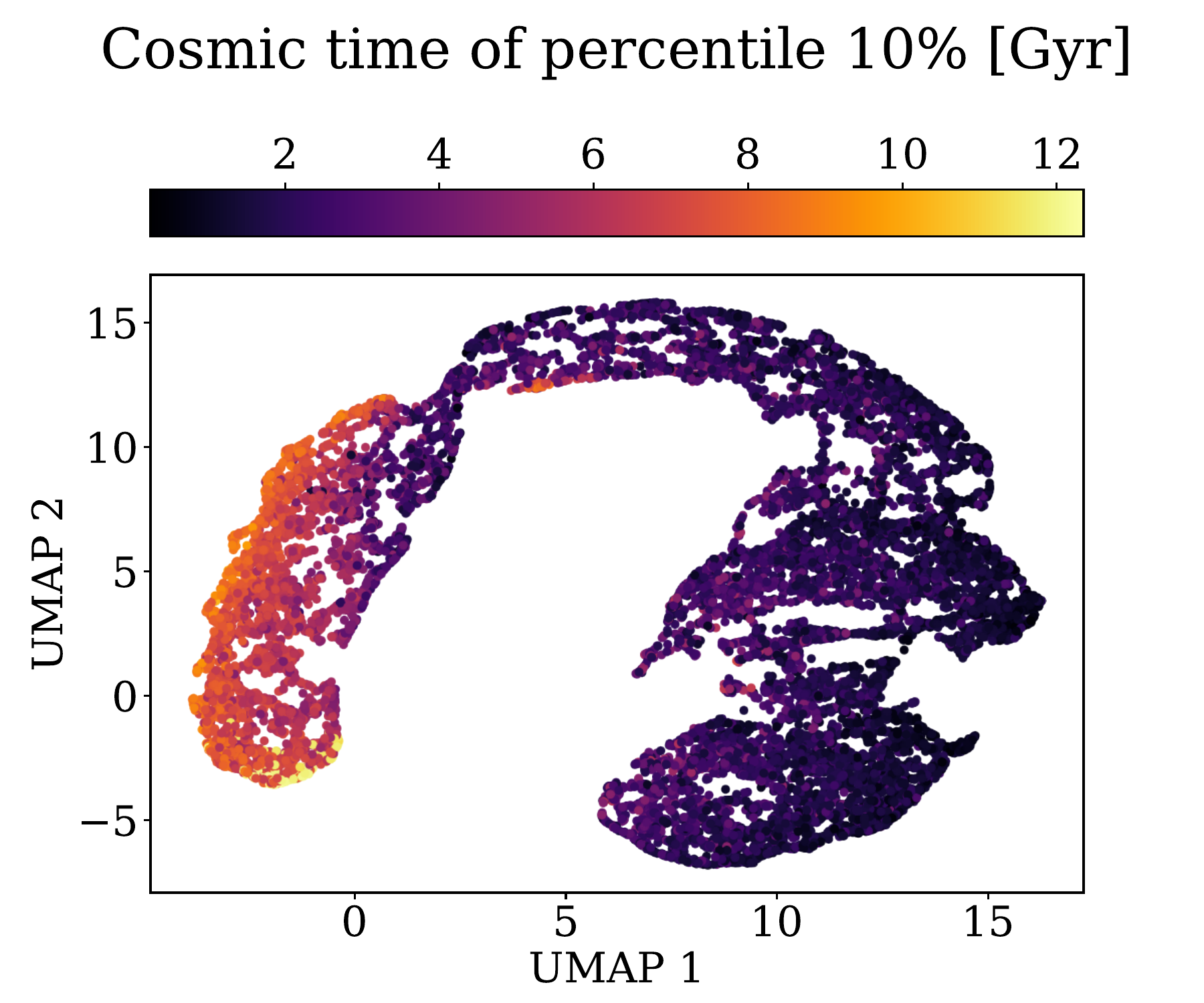}
    \includegraphics[width=0.33\textwidth]{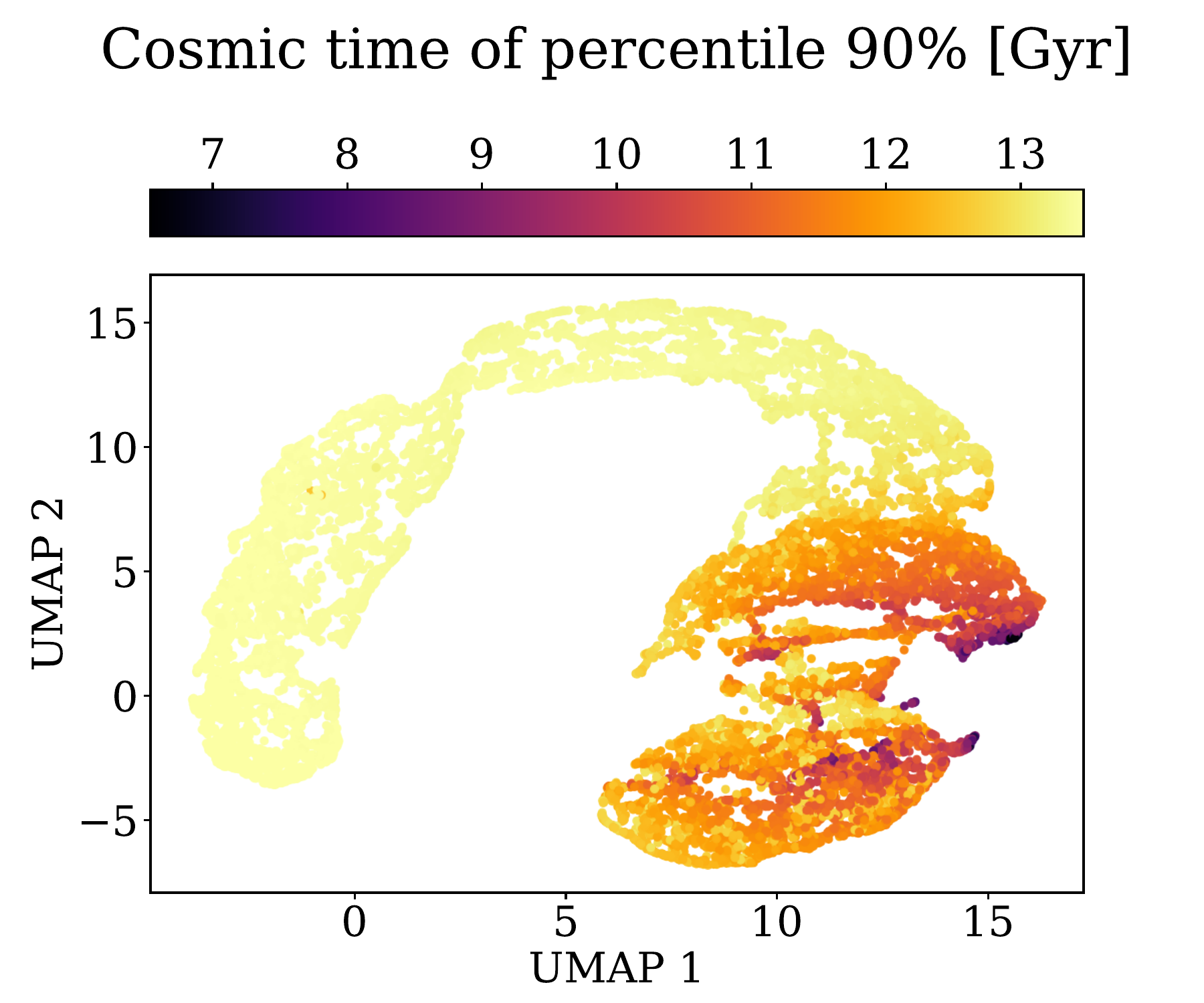}
    \includegraphics[width=0.33\textwidth]{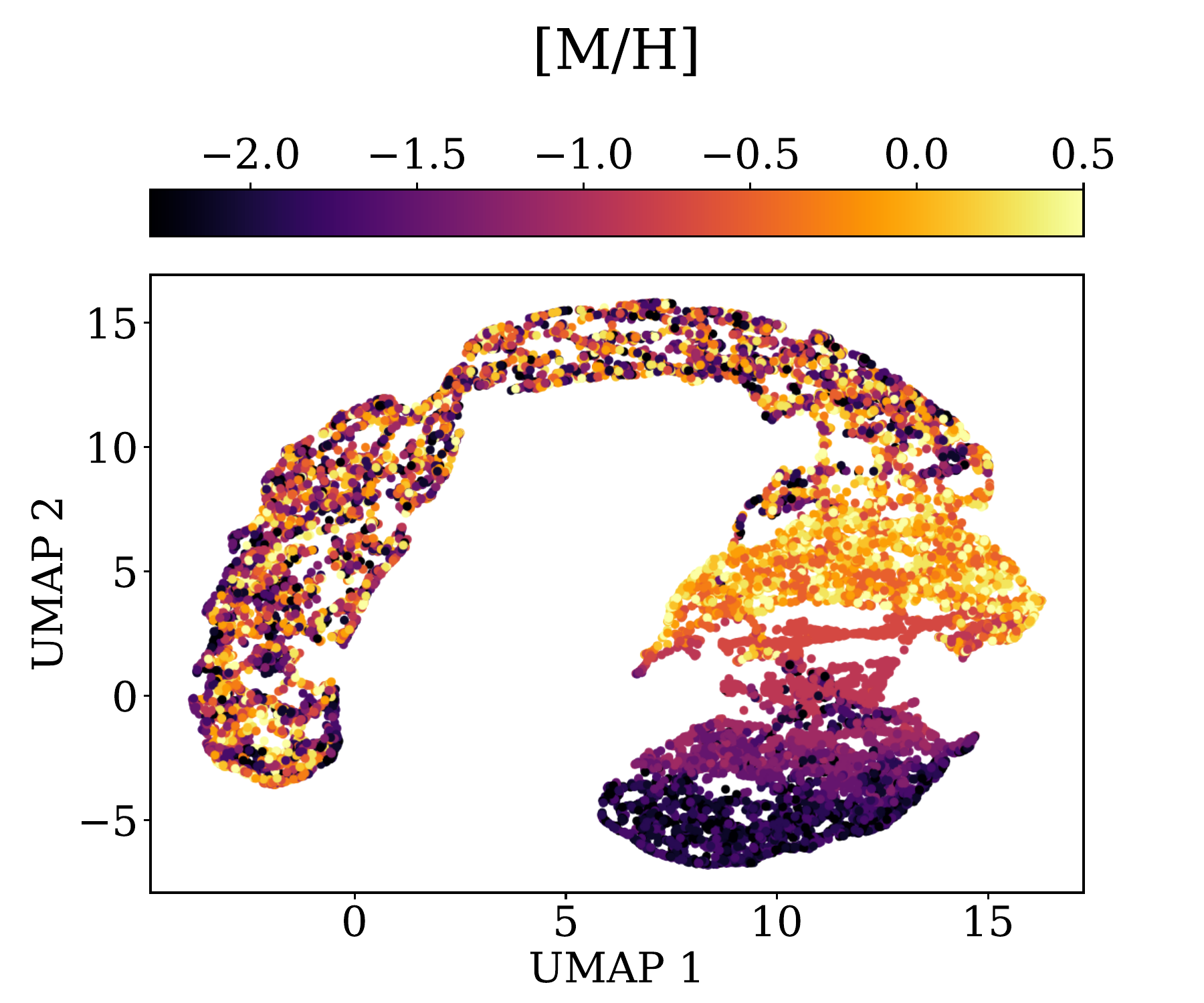}
    \includegraphics[width=0.33\textwidth]{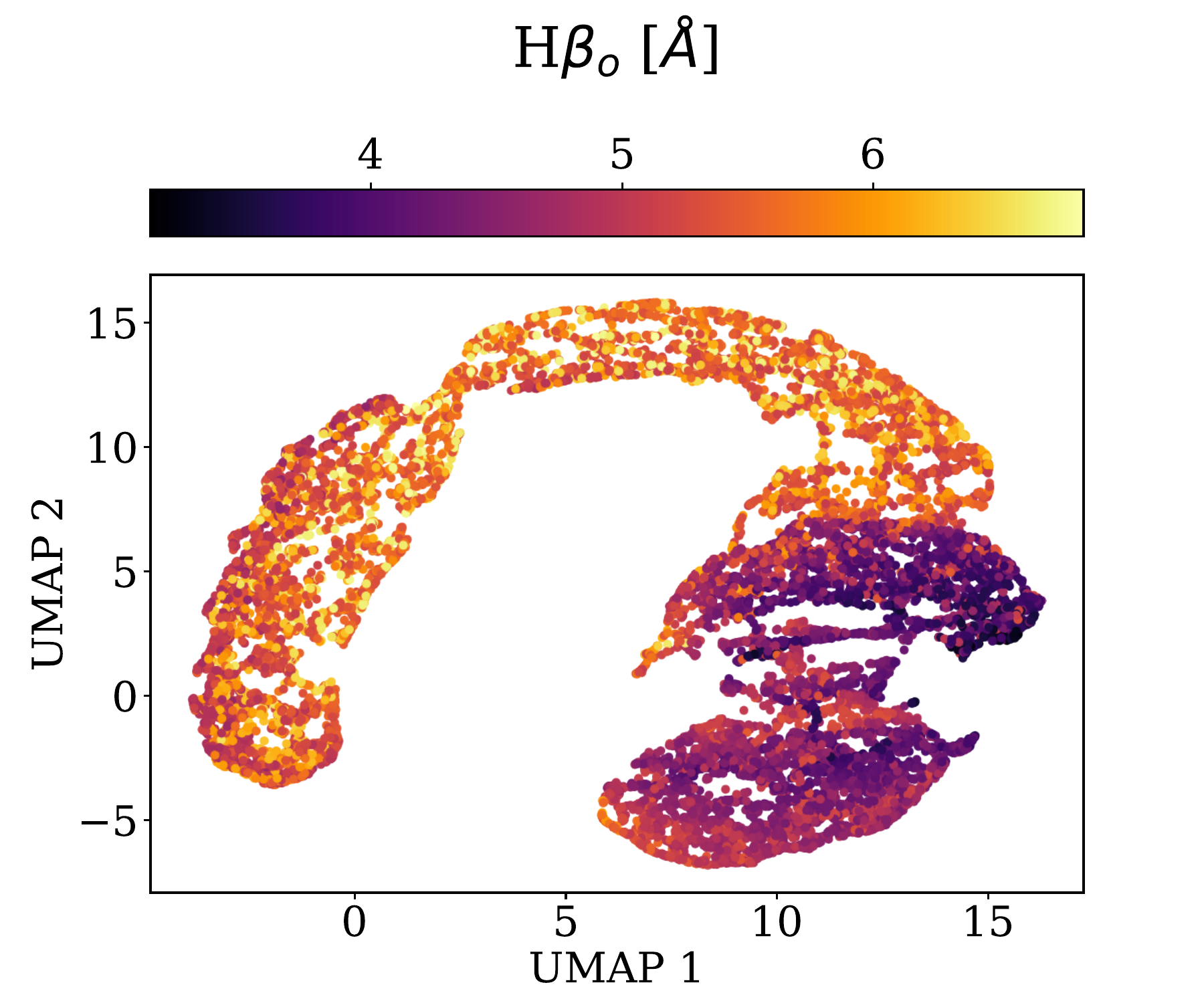}
     \includegraphics[width=0.33\textwidth]{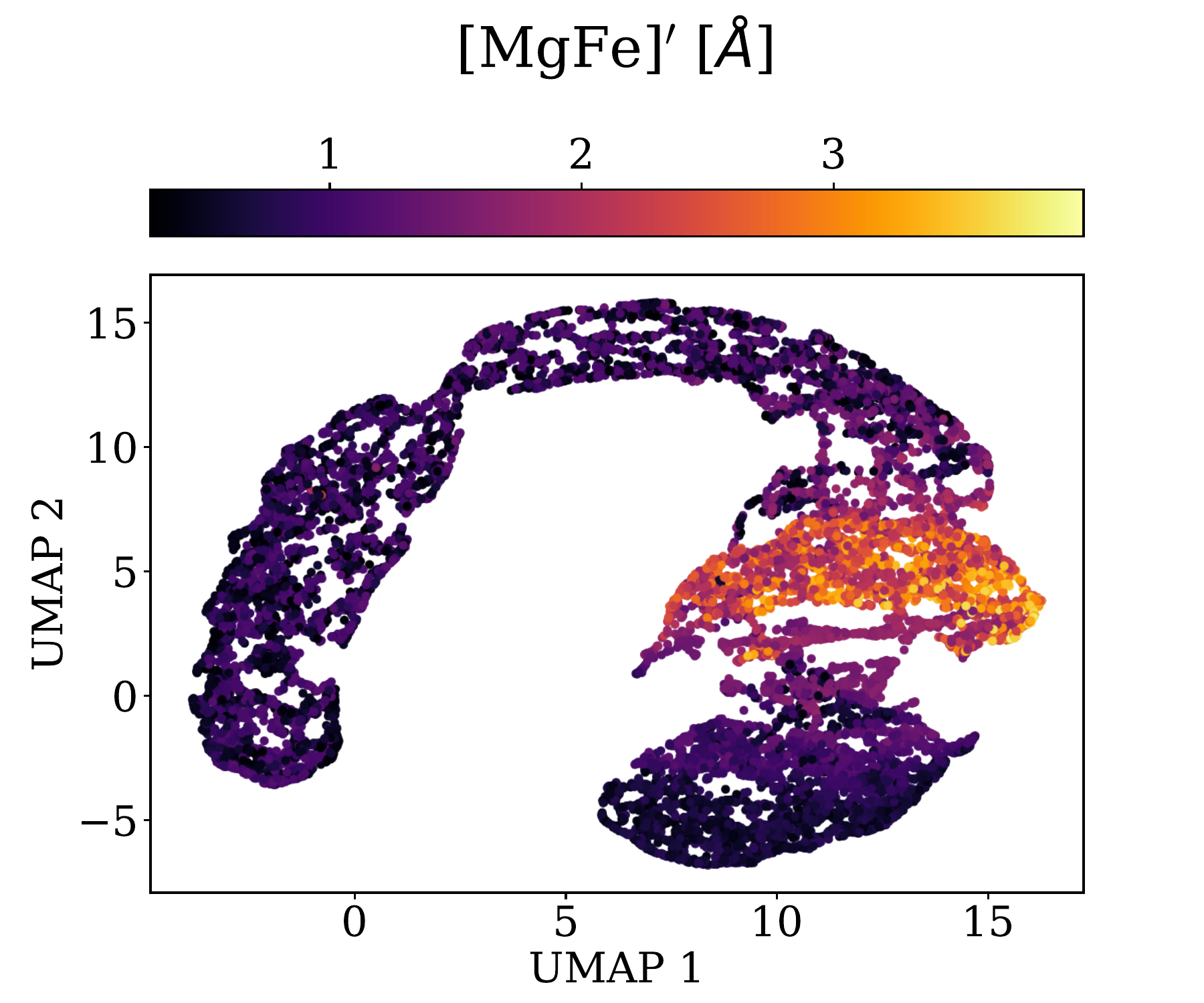}

    \caption{UMAP embedding for the latent representations. Each dot corresponds to a synthetic galaxy from the test sample. The closer two galaxies are on the map, the more similar are their latent representations. Different colour maps are set according to the cosmic time at which $10\%$ (top left) and $90\%$ (top middle) of the total stellar mass is reached, the $[\rm{M/H}]$ (top right), and the two line indices  $\rm H{\beta_{o}}$ (lower left) and $[\rm{MgFe}]^{\prime}$ (lower right). Clear trends can be observed, demonstrating that the information encoded in the latent vectors is optimal for recovering the SFHs and metallicities.}
    \label{umap}
    
\end{minipage}
\end{figure}

\begin{minipage}[t][0.4\textheight][t]{\textwidth}
\begin{multicols}{2}

We did a sanity check to verify whether the encoder is well trained and performs a reliable summary statistics of the spectra and the physical properties. A uniform manifold approximation and projection (UMAP) of the latent vectors is shown in Fig.~\ref{umap}, only for visualisation purposes. It uses a non-linear dimensionality reduction technique that produces two-dimensional maps topologically equivalent to the latent vectors of $16$ components \citep[see][]{umap}. The shape of the UMAP, where each dot corresponds to a synthetic galaxy from the test sample, can be intuitively explained according to the physical quantities we want to recover, demonstrating an optimal encoding of the spectral information. The map is shown using five different colour maps: the value of the cosmic time for the $10\%$ and $90\%$ stellar mass percentiles, the metallicity, and the line indices  $\rm H{\beta_{o}}$ \citep{cervantes09} and $[\rm{MgFe}]^{\prime}$  \citep{thomas03}.\\

First, the percentile $10\%$ plot show how galaxies clustered on the left started forming their stars late (large cosmic time for the percentile $10\%$), while on the centre and right of the UMAP there are located galaxies that started forming stars early (small cosmic time for the percentile $10\%$). Focusing now on the percentile $90\%$ and $\rm H{\beta_{o}}$ plots,  we observe that last episodes of star formation take place at larger values of the cosmic time for the left and upper clusters, while the galaxies in the lower right have stopped forming stars earlier. In short, if we divide the UMAP into left, upper, and right zones, we see that the galaxies on the left have a short and recent star formation, those on the top have extended and recent star formation, and those on the right have old star formation.  Furthermore, in the right region there are two clusters with no difference in age, but in metallicity, splitting galaxies without recent star formation into two groups: the upper one, metal-rich, and the lower one, metal-poor (see the metallicity and $[\rm{MgFe}]^{\prime}$ plots).  Finally, we highlight that in the galaxies with more recent star formation, there is no clear distinction in metallicity. The complexity of measuring metallicity in young populations is already known \citep{Conroy_2013} and is based on stellar physics, as the high temperatures of the atmospheres of massive stars cause the metallic lines to be very faint, and the parameter that fundamentally governs the spectra is age. The network, without any constraints in the input, naturally recovers the underlying physics of spectral fitting, as well as the fundamental role of the  absorption lines used in measurements of the indices \citep{Vazdekis_2010}.

\end{multicols}
\end{minipage}

\end{appendix}

\end{document}